\pdfoutput=1
\documentclass[journal]{vgtc}                % final (journal style)
\usepackage[utf8]{inputenc}

\ifpdf%                                % if we use pdflatex
  \pdfoutput=1\relax                   % create PDFs from pdfLaTeX
  \usepackage{graphicx}                % allow us to embed graphics files
  \DeclareGraphicsExtensions{.pdf,.png,.jpg,.jpeg} % for pdflatex we expect .pdf, .png, or .jpg files
\else%                                 % else we use pure latex
  \usepackage{graphicx}                % allow us to embed graphics files
  \DeclareGraphicsExtensions{.eps}     % for pure latex we expect eps files
\fi%

\graphicspath{{figures/}{pictures/}{images/}{./}} % where to search for the images

\usepackage{microtype}                 % use micro-typography (slightly more compact, better to read)
\PassOptionsToPackage{warn}{textcomp}  % to address font issues with \textrightarrow
\usepackage{textcomp}                  % use better special symbols
\usepackage{mathptmx}                  % use matching math font
\usepackage{times}                     % we use Times as the main font
\usepackage{cite}                      % needed to automatically sort the references
\usepackage{tabu}                      % only used for the table example
\usepackage{booktabs}                  % only used for the table example

\newcommand\systemname{ClinicalPath}

\newcommand{\myParagrapho}[1]{\smallskip\noindent\textbf{#1}\xspace}

\newcommand*{\img}[1]{%
    \raisebox{-.3\baselineskip}{%
        \includegraphics[
        height=0.8\baselineskip,
        width=0.8\baselineskip,
        keepaspectratio,
        ]{#1}%
    }%
}

\onlineid{0}

\vgtccategory{Research}
\vgtcpapertype{algorithm/technique}

\usepackage{cellspace}

\usepackage{ifpdf}

\ifpdf%                                % if we use pdflatex
  \pdfoutput=1\relax                   % create PDFs from pdfLaTeX
  \usepackage{graphicx}                % allow us to embed graphics files
  \DeclareGraphicsExtensions{.pdf,.png,.jpg,.jpeg} % for pdflatex we expect .pdf, .png, or .jpg files
\else%                                 % else we use pure latex
  \usepackage{graphicx}                % allow us to embed graphics files
  \DeclareGraphicsExtensions{.eps}     % for pure latex we expect eps files
\fi%

\usepackage{tabularx}

\graphicspath{{figures/}{pictures/}{images/}{./}}

\usepackage{microtype}         
\PassOptionsToPackage{warn}{textcomp} 
\usepackage{textcomp}                 
\usepackage{mathptmx}                
\usepackage{times}                    
       
\usepackage{cite}                   
\usepackage{tabu}                   
\usepackage{booktabs}               
\usepackage[normalem]{ulem} 
\usepackage{xspace}
\usepackage{amsfonts}
\usepackage{amsmath}
\usepackage{multirow}  % For table

\usepackage[utf8]{inputenc}
\usepackage[T1]{fontenc}
\usepackage{balance}
\usepackage[hyphens]{url}
\usepackage{tabularx}

\usepackage{wrapfig,epsf}

\usepackage{xcolor}

\usepackage[hidelinks]{hyperref}

\newcommand{\ie}{\emph{i.e.}\xspace}
\newcommand{\eg}{\emph{e.g.}\xspace}

\newcommand{\figref}[1]{\hyperref[#1]{\textbf{Figure~\ref*{#1}}}}
\newcommand{\figsref}[1]{\hyperref[#1]{Figures~\ref*{#1}}}
\newcommand{\tabref}[1]{\hyperref[#1]{Tab.~\ref*{#1}}}
\newcommand{\secref}[1]{\hyperref[#1]{Sec.~\ref*{#1}}}
\newcommand{\algref}[1]{\hyperref[#1]{Alg.~\ref{#1}}}

\title{\systemname: a Visualization tool to Improve the Evaluation of Electronic Health Records in Clinical Decision-Making}

\author{Claudio~D.~G.~Linhares*,
        Daniel~M.~Lima,
        Jean~R.~Ponciano, 
        Mauro~M.~Olivatto,
        Marco~A.~Gutierrez, \\
        Jorge~Poco, 
        Caetano~Traina~Jr.,
        and Agma~J.~M.~Traina}

\authorfooter{
\item *C. Linhares is the corresponding author.
\item C. Linhares, D. Lima, C. Traina Jr., and A. Traina are with Institute of Mathematics and Computer Sciences, University of S\~ao Paulo, S\~ao Carlos, Brazil.
\protect\\
E-mails: \{claudiodgl, danielm\}@usp.br, \{caetano, agma\}@icmc.usp.br.
\item D. Lima, M. Gutierrez are with Laboratorio de Informatica Biomedica, Instituto do Coracao, Hospital das Clinicas HCFMUSP, Faculdade de Medicina, Universidade de Sao Paulo, Sao Paulo, SP, Brazil. 
\protect\\
E-mails: \{daniel.lima, marco.gutierrez\}@incor.usp.br.
\item J. Ponciano and J. Poco are with School of Applied Mathematics, Getulio Vargas Foundation, Rio de Janeiro, Brazil. 
\protect\\
E-mails: \{jean.ponciano, jorge.poco\}@fgv.br.
\item M. Olivatto is with Graduate Course in Medicine, Federal University of Fronteira Sul, Chapec\'o, Brazil. \protect\\
E-mail: mauro\_olivatto@hotmail.com.

}

\abstract{
Physicians work at a very tight schedule and need decision-making support tools to help on improving and doing their work in a timely and dependable manner. Examining piles of sheets with test results and using systems with little visualization support to provide diagnostics is daunting, but that is still the usual way for the physicians' daily procedure, especially in developing countries. Electronic Health Records systems have been designed to keep the patients' history and reduce the time spent analyzing the patient's data. However, better tools to support decision-making are still needed. In this paper, we propose \systemname, a visualization tool for users to track a patient's clinical path through a series of tests and data, which can aid in treatments and diagnoses. 
Our proposal is focused on patient's data analysis, presenting the test results and clinical history longitudinally. Both the visualization design and the system functionality were developed in close collaboration with experts in the medical domain to ensure a right fit of the technical solutions and the real needs of the professionals. We validated the proposed visualization based on case studies and user assessments through tasks based on the physician's daily activities. Our results show that our proposed system improves the physicians' experience in decision-making tasks, made with more confidence and better usage of the physicians' time, allowing them to take other needed care for the patients.
}

\vgtcinsertpkg

\begin{document}


\maketitle

\section{Introduction}
\label{sec:introduction}

The day-to-day activities of health professionals working in medical centers and hospitals are challenging. Physicians are frequently responsible for delicate and high-impact decisions over a patient's condition that must be taken quickly and precisely. Adequate tools can save professionals time, improve diagnostic procedures and improve patients' quality of life. In a long-term illness or a critically ill patient case, examining and comparing large amounts of data can be disorienting and tiresome, creating misleading impressions. All of these problems have been exacerbated by the COVID-19 pandemic, which has imposed long work shifts and extra time dedication on health specialists, resulting in less time for interpretations of the patient's test history. Offering good health services is one of the significant challenges in developing countries, such as Brazil~\cite{brazil_EHR1} and others. Although recent studies show that Electronic Health Records (EHR) are being used in those countries, they are concentrated in capitals and better-developed areas~\cite{brazil_EHR3}. 

EHR systems were proposed to store the patients' clinical history, support clinical decisions, and speed up patient analyses. These systems can improve the physicians' perception of the patient's data and enhance diagnostic accuracy~\cite{survey_EHR1}. EHRs can support the tasks of interpreting, predicting, and monitoring~\cite{survey_EHR2}, and often employ Information Visualization techniques. Many systems were proposed focusing on information from one or multiple patients, depending on the intended analysis~\cite{survey_EHR2}. Nevertheless, tools with better visual representations and scalability complying with the physician's needs are still necessary to support tasks that identify diagnostic hypotheses, treatments or take changes in patient's tests evolution into account.

In this paper, we propose the \systemname~system, an interactive and freely available\footnote{\url{github.com/claudiodgl/ClinicalPath}} visualization tool that assists physicians in their daily routine. This system incorporates a longitudinal map of the patient information evolution and presents the patient tests' results and clinical history using symbols and other visual metaphors. An initial idea of the \systemname~was proposed in the \textit{Interoperable Covid Visualizer (I-CovidVis)} tool~\cite{icovidvis_cbms}, which served as an initial baseline.
To test our system, we selected cases from the FAPESP COVID-19 Data Sharing/BR repository~\cite{covid-datasharingbr2020,artigobaseFapesp1}, which contains data from five healthcare institutions in Brazil, with more than 30 million laboratory test results from patients with or suspected of COVID-19.

The main contributions of this paper are summarized as follows:

\begin{itemize}

\item An effective EHR visualization technique that allows users to track a patient's clinical path and enables quick test results analysis in a user-specified timeline.

\item \systemname, an interactive and freely available software that implements the proposed visualization and many interactive tools. The system was developed in close collaboration with physicians, including domain experts in the clinical medical modality.

\item A robust analysis of the efficacy, effectiveness, intuitiveness, and usability through case studies and user evaluation.

\end{itemize}

\section{Related Work and Concepts}
\label{sec:related_work}
EHR systems are used to analyze from a patient's history to the effect of treatments, thus greatly benefiting clinical decisions~\cite{survey_EHR2}.
EHR systems can be improved by Information Visualization techniques, such as appropriate layouts, visual encoding, and interactive tools. Recent surveys identified that EHRs focus on two types of systems: population-based and single-patient tools~\cite{survey_EHR2}. An EHR comprises three main tasks: \textit{data interpretation}, which is the activity of detecting patterns in the patient records; \textit{prediction}, which estimates the odds of patient's outcome or future diseases; and \textit{monitoring}, which investigates sequences of aggregated events~\cite{survey_EHR2}. Given the context in which \systemname~is employed, the rest of this section discusses single-patient and population-based visualization systems whose layouts are partially or totally dedicated to analyzing test results or clinical history.

\textit{Population-based tools} focus on general behaviors of simultaneously visualized patient records. Two of the most popular tools in this category are LifeLines2~\cite{LifeLines2}, a system focused on summarizing patient's records over time and detecting/comparing patterns over groups of records; and VISITORS~\cite{VISITORS}, which aggregates raw data for several patients displayed over time, for exploratory tasks on clinical data.
Both systems are suited for interpreting and monitoring tasks~\cite{survey_EHR2}. There are also systems with other goals. Some of them are focused, for example, on the evaluation of adherence to a treatment plan~\cite{5742371} or patient cohort~\cite{informatics7020017}, \ie, groups of individuals affected by common diseases, treatments, and comorbidities.

Although population-based tools are currently the most popular~\cite{survey_EHR2}, they usually focus on general tasks, such as insights considering a set of patients to find similarities or trends in test results. Therefore, population-based EHR tools are not meant for individual diagnostic hypotheses or patients' clinical analyses.

On the other hand, single-patient tools focus on local behaviors of visualizing a patient record, such as comparing the test results and understanding their variation over time. Popular tools include MIVA~\cite{MIVA}, HARVEST~\cite{HARVEST}, and Dabek et al.'s~\cite{8387501}. MIVA is a tool that displays data from bedside biometric devices and health care in Intensive Care Units (ICU). The visualization shows the biometric information as rows and the timestamps as columns. Test values are displayed using point plots, keeping track of the changes over time. Although MIVA has a focus similar to ours, its visualization is based on one scatter plot for each test and requires a lot of screen space to be displayed, thus hampering visual scalability. This proposal is unfeasible in our context, as it would require many user interaction steps (such as filter tuning) to present an equivalent amount of information. HARVEST focuses on summarizing the patient's data using textual visualization, word clouds, and timeline representations to assist physicians in interactive searches to improve patient care~\cite{HARVEST}. Although some visual encodings aim at helping domain experts, they rely on textual information, which requires more time for experts to perform the required clinical tasks as they have to read potentially lengthy textual descriptions instead of visualizing test results representations. At last, Dabek et al.~\cite{8387501} present a system that focuses on summarization and simplification. It offers several visual encodings for diagnoses, medications, and patient history. However, the system still lacks evaluation with domain specialists to demonstrate its usefulness and usability.

There are other systems also focused on single-patient analysis, such as CareVis~\cite{CareVis}, Steinhauer et al.'s~\cite{Steinhauer}, VisuExplore~\cite{VisuExplore}, and Midgaard~\cite{Midgaard}.  
CareVis uses a scatter plot for each test over time, resulting in unused screen space. VisuExplore focuses on chronic diseases through interactive visualization techniques for single and multiple patients. The system has multiple views for different data types and tasks, with line charts for test results, bar charts, glyphs, and timelines for other views. Midgaard relies on interactive visualizations, specifically in timeline navigation, to create a temporal distortion that improves the exploration of diagrams and line charts. Steinhauer et al.'s system presents a visualization of image data, such as computed tomography and magnetic resonance imaging, using information architecture to organize the relevant clinical information, focused on a medical specialization of dermatologists and oncologists. The visualization is divided into two main parts: (i) therapy data and (ii) test results. The therapy data is composed of different symbols to represent clinical stages, and the tests are represented by line charts, also presenting the same previously mentioned scalability problems.

Another single-patient system is MTSA~\cite{MTSA}, which uses animated radial parallel coordinates to visualize multivariate data, such as laboratory and physiological data. The MTSA's approach differs from others because it relies on animation rather than a timeline-based visualization. Animations may impair the user's mental map preservation. Furthermore, this radial approach limits the number of tests displayed on the screen due to visual cluttering and makes it harder to compare and examine multiple results simultaneously. At last, there are single-patient commercial systems that provide timeline visualizations of clinical events, such as the ICUdata~\cite{ICUData}, which also focus on line charts to represent test results, having the previously mentioned problems.

Our focus is to analyze test results and clinical history composed of several records to build a diagnostic hypothesis.
Existing systems fulfill part of the medical staff requirements~\cite{glicksberg_patientexplorer_2019, othersystems1,othersystems2, othersystems3}, and our system provides visualizations for other features relevant for data analysis, developed to fulfill the requirements provided by the domain experts (details in Section~\ref{sec:methodology}). We also pursued several guidelines suitable for EHRs, such as those described in~\cite{MIVA}: \textit{"Keep display simple and free of clutter"}, \textit{"Provide orientation clues"}, \textit{"Include appropriate graphics that support and clarify data"}, \textit{"Provide comparisons to references and normal limits"}, and others. We keep the timeline approach, as it is the main layout provided by almost every EHR, and professionals are used to it. Thus, it was used since the initial  \systemname~interaction~\cite{icovidvis_cbms}, and extended it with several improvements. 

Table~\ref{tab:comparison} shows a comparison between \systemname~and several EHR systems according to some well-established aspects~\cite{survey_EHR1,survey_EHR2} related to the goal of this work. Inspired by~\cite{6415893}, we compared them using the corresponding papers' descriptions, software images, case studies, comprehensive surveys (\eg,~\cite{survey_EHR1,survey_EHR2}), and external videos, when available.

\begin{figure*}[!t]
	\includegraphics[width=1\linewidth]{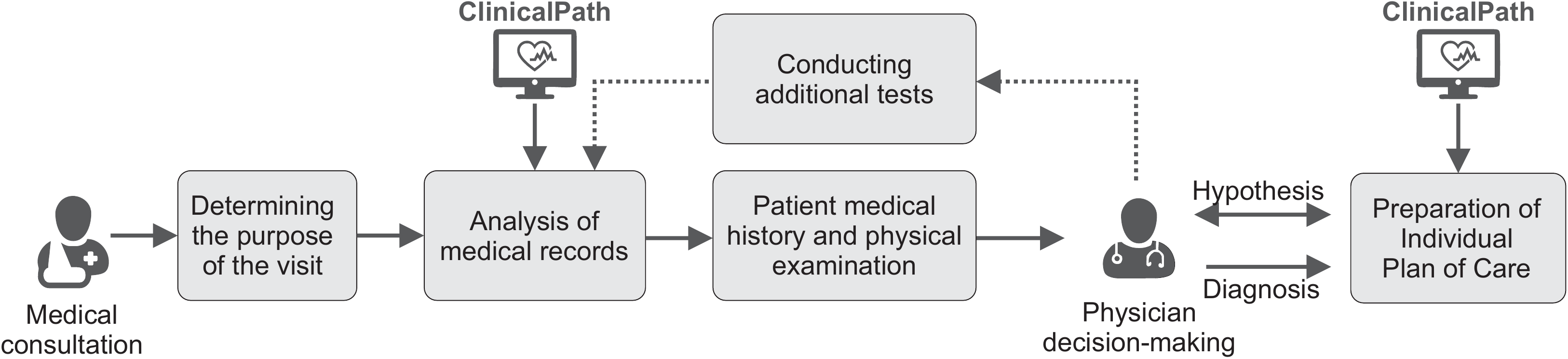}
	\centering
	\caption{General description of the clinical workflow of a physician in a regular medical consultation, and the importance of the \systemname~system to help the physician in the decision-making process.
	}
	\label{fig:workflow_ClinicalPath}
\end{figure*}

The first three aspects are selecting (1), grouping (2), and filtering (3). For comparison purposes, these aspects are related to tests and time intervals. We also consider the classification of test results (4), clinical history and outcome (5), and whether the system was evaluated by domain experts (6). At last, we consider aspects related to visual scalability (7), \ie, the system's capability to comprise much information without requiring much screen space. According to the raised aspects, VISITORS and LifeLines2 are the systems most similar to ours. However, they do not meet \systemname~objectives, as they are population-based tools. 

\begin{table}[ht]
\centering
\caption{Comparison of EHR systems according to seven different aspects: (1) selecting; (2) grouping; (3) filtering; (4) test result classification; (5) clinical history and outcome; (6) user evaluation; (7) scalability of tests over time. }
\label{tab:comparison}
\resizebox{0.5\textwidth}{!}{%
\begin{tabular}{lccccccc}
\hline
             & 1 & 2 & 3 & 4 & 5 & 6 & 7 \\ \hline
VISITORS~\cite{VISITORS}     & $\checkmark$ & $\checkmark$ & $\checkmark$  & $\checkmark$ &   & $\checkmark$ &   \\
LifeLines2~\cite{LifeLines2}   & $\checkmark$ & $\checkmark$ & $\checkmark$  & $\checkmark$  &   & $\checkmark$ &   \\
MIVA~\cite{MIVA}         & $\checkmark$ &   & $\checkmark$  &   &   & $\checkmark$ &   \\
Dabek et al.'s~\cite{8387501} & $\checkmark$ &   & $\checkmark$ &  $\checkmark$ &   &   &  $\checkmark$  \\
HARVEST~\cite{HARVEST}      &   & $\checkmark$ & $\checkmark$ &    & $\checkmark$ & $\checkmark$ &   \\
\color{black}{CareVis~\cite{CareVis}} & $\checkmark$ & $\checkmark$ &  &  &  & $\checkmark$ &  \\
\color{black}{VisuExplore~\cite{VisuExplore}} & $\checkmark$ & $\checkmark$ & $\checkmark$ &   &   & $\checkmark$ &  \\
\color{black}{Midgaard~\cite{Midgaard}} & $\checkmark$ &   & $\checkmark$ &   &  &  & $\checkmark$ \\
\color{black}{Steinhauer et al.'s~\cite{Steinhauer}} & $\checkmark$ &   &   & $\checkmark$  & $\checkmark$ & $\checkmark$ &  \\
\color{black}{MTSA~\cite{MTSA}} & $\checkmark$ &   & $\checkmark$ & $\checkmark$  &  & $\checkmark$ &  \\
ClinicalPath (Our) & $\checkmark$ & $\checkmark$ & $\checkmark$ & $\checkmark$  & $\checkmark$ & $\checkmark$ & $\checkmark$ \\ \hline
\end{tabular}%
}
\end{table}

\section{Requirements and Tasks}
\label{sec:methodology}

Our system was designed according to the requirements posed by the domain experts to meet their needs and suggestions. The requirements were identified and formulated based on discussions in weekly meetings among the authors of this paper, who are experts in visual analytics, biomedical informatics, and the medical domain. We raised requirements targeting users' potential interests (led by biomedical informatics and medical experts), and then we designed visual analytics tasks (led by visual analytics experts). At last, we explained a clinical workflow for a medical consultation, which shows the physician's decision-making process using the \systemname~system.

\subsection{System Requirements}

We identify and highlight the key needs of domain experts in order to enumerate the main requirements for a system, which we organized into six statements regarding what the physicians want to know:

\myParagrapho{S1.} How the test results change over time.

\myParagrapho{S2.} What tests are performed on a specific day or period of time.

\myParagrapho{S3.} Compare tests’ results of a patient over time (patient history).

\myParagrapho{S4.} Which departments of the facility is/was the patient located in.

\myParagrapho{S5.} If and What part of the patient's event history is available.

\myParagrapho{S6.} Analyze tests results according to target specialties, such as endocrine, cardio, renal ones.

\subsection{Design Tasks}

We designed the following visualization tasks to meet the raised requirements, providing a meaningful and valuable tool for the target user. We categorized the tasks according to the typology of abstract visualization tasks~\cite{6634168}. This categorization comprises \textit{higher-level goals} (why), such as discover, search, and comparison; \textit{lower-level goals} (how), such as filter, arrange, navigate; and \textit{user-specific tasks} (what), such as inputs and outputs.

\myParagrapho{T1 -- Exhibit and classify test results (how):} The visualization should arrange how the tests' results appear, according to chronological order. It should also highlight how close or far they are to their corresponding reference values (S1, S3).

\myParagrapho{T2 -- Select tests and results according to date (how):} The visualization should enable the navigation in the layout, allowing the selection of different tests and their results for a single day or a time interval (S2).

\myParagrapho{T3 -- Filter only dates with tests performed (how):} The visualization should enable filtering by dates to facilitate the selection (S2) and comparison (S3) of the test results over time.

\myParagrapho{T4 -- Compare a set of test results over time (why):} The visualization should allow a quick and intuitive comparison of different and related test results for different dates (S3).

\myParagrapho{T5 -- Map clinical history and outcomes (what):} The visualization should allow understanding of the patient's events, including those related to the patient care, and the corresponding outcome, when applicable (S4, S5).

\myParagrapho{T6 -- Group and order tests (how):} The visualization should identify different groups of tests and define how to order them, making it easy to compare results (S3) and guide the analyses regarding different specialties (S6).

\subsection{Clinical Workflow}

Based on previous studies~\cite{Szelagowski2021}, we illustrated in Fig.~\ref{fig:workflow_ClinicalPath} a regular clinical workflow to demonstrate the steps of the physician in a medical consultation. In the first step, the physician collects key patient data, such as age, sex, profession, patient's address, and weight, and understands the purpose of the visit. After, the physician checks previous medical records of the patient to match the described conditions, which can directly be consulted in the \systemname~system. Then, the physician performs a physical examination and checks the patient medical history, such as the family history of diseases (\eg, genetic diseases), medications of continuous use, history of previous surgeries, history of hospitalizations, allergies, and others. In the decision-making phase, the physician can create a hypothesis, which leads to preparing an immediate care plan to alleviate immediate problems and consequently perform additional tests to confirm or discard the hypothesis. Also, if the physician has enough information, the diagnosis can be indicated, and a therapy to treat the patient's problem is started. At future patient returns, the physician can execute the clinical workflow again to verify whether the therapy was successful. Notice that, in the step to prepare the individual plan of care, the \systemname~also plays a crucial role in helping the physician analyze the test results to indicate appropriate doses of medications or recommended procedures.

\section{Data Set}
\label{sec:dataset}

The \textit{FAPESP COVID-19 Data Sharing/BR} Repository~\cite{covid-datasharingbr2020} contains anonymous patient information related to COVID-19, such as clinical tests, outcome, and demographic information (age, sex, birthplace)~\cite{artigobaseFapesp1}. The database has been updated frequently with new Brazilian institutions and data. The  following discussion is based on the data available in January 2021, which contains data from five institutions: Albert Einstein Hospital, Fleury laboratories, S\'irio-Liban\^es Hospital, Benefic\^encia Portuguesa/SP Hospital, and Hospital das Clinicas referred here as HF1, HF2, HF3, HF4, and HF5, respectively.
Each laboratory result has the test identification, the analyte (\ie, the material or chemical constituent measured), the obtained and reference values, and the unit of measure. For the sake of simplicity, we refer to both the test and analyte as a test in this paper.
Two institutions (HF3 and HF4) also provided the patient outcome. Table \ref{tab:cardinality} shows the cardinality of each type of entity from the five institutions.

\begin{table}[ht]
\centering
\caption{Number of patient tests, measurements and outcomes, and the total amount in the five institutions. HF1, HF2, and HF5 do not provide patient outcomes.
    }
\label{tab:cardinality}
\resizebox{0.49\textwidth}{!}{%
\begin{tabular}{lcccccc}
\hline
\textbf{Entity} & \textbf{HF1} & \textbf{HF2} & \textbf{HF3} & \textbf{HF4} & \textbf{HF5} & \textbf{Total} \\ \hline
patients        & 79K          & 470K         & 4K           & 39K          & 3K           & 595K           \\
results         & 3,415K       & 19,274K      & 631K         & 6,329K       & 2,498K       & 32,147K        \\
outcomes        & -            & -            & 16K          & 217K         & -            & 233K           \\ \hline
test types          & 63           & 830          & 537          & 716          & 435          & 2,581          \\
analytes types        & 126          & 1K           & 731          & 1K           & 896          & 3,753          \\ \hline
\end{tabular}%
}
\end{table}

Previous studies using this data set highlighted problems with the data and demonstrated the importance of data cleaning and normalization steps to enable real-world analyses~\cite{artigobaseFapesp2}. Different objectives have guided studies that consider this data set. One such study focused on finding relations involving patients' gender, age, and immunological response to COVID-19\,---\, which included the discovery that levels of markers of inflammation in critically ill diagnosed COVID patients were extremely high and varied by sex and age~\cite{artigobaseFapesp3}. In another study, the authors analyzed the similarity and correlation of tests from this data set with another data set of chest X-ray data images~\cite{correlation_FAPESP}. They have identified a strong correlation of pneumonia with positive COVID tests for male patients. Although existing studies also focused on EHR systems, they concentrated on global and general findings involving tests and patients rather than local analysis of diagnostic hypotheses for each patient.

\section{Data Pre-processing}
\label{sec:preprocessing}

\myParagrapho{Data Cleaning and Normalization.} Each healthcare institution uses different types of tests, analytes (often with diverse vocabularies and spellings), and units of measurement (mostly following the International System of Units, but at different scales). The raw data contains noise, character typos, extraneous punctuation, and missing data. To perform the analysis and create an equivalence between the institutions to make data interoperable, we had to normalize and clean all the raw data, including the test and analyte names, adjust scales, and unify representations (as described in~\cite{icovidvis_cbms}). Also, we choose to discard tests with noise and missing data, such as incomprehensible test names, incomplete values (test results, dates, scales), and so on.

After the cleaning and normalization steps, the test types were reduced from 2,581 to 73 unique values. Although there was a considerable reduction in the number of test types, the normalization, in this case, decreases the number of variables but does not necessarily discard data. For instance, since this data set deals with five different institutions, the data had a lot of different spelling variations for the same test, which was the case for several COVID tests. Moreover, we discarded very rare tests for which there was not enough data to make distribution estimates for the abnormality ranges, which was one of the goals of this work, as will be detailed as follows. Also, the original data with more than 32 million result tests was reduced to 10 million. Since we focus on analyzing single patients from different institutions, these cleaning and normalization steps are at the core of the interoperable data sharing among the institutions but have little or no impact on our analysis, as will be discussed in Section~\ref{sec:discussionlimitations}. The list of tests, analyses, and other descriptions is available in the supplemental material.

\begin{figure}[!b]
	\includegraphics[width=1\linewidth]{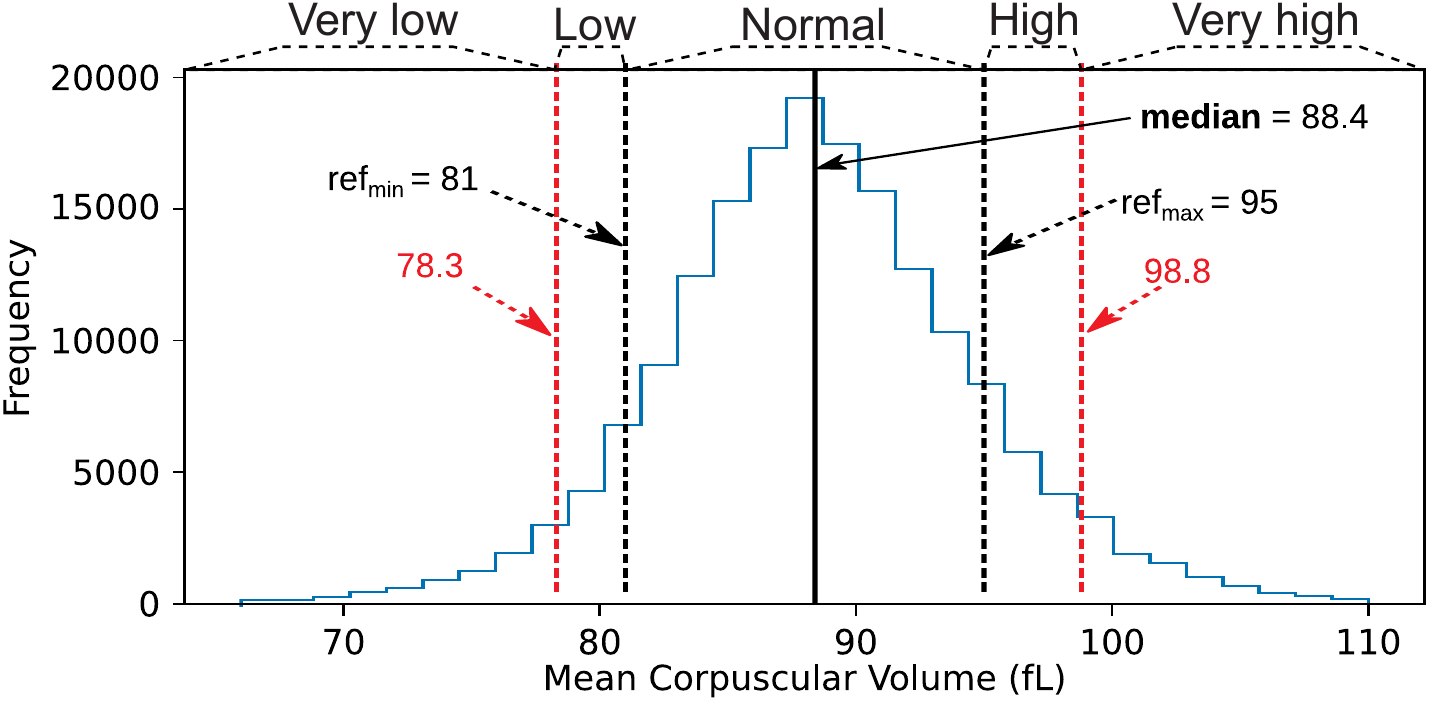}
	\centering
	\caption{Example of the calculation of the five categories using the reference values, and the medians in low/high subsets. 
	}
	\label{fig:reference_value}
\end{figure}

\begin{figure*}[ht]
	\includegraphics[width=\linewidth]{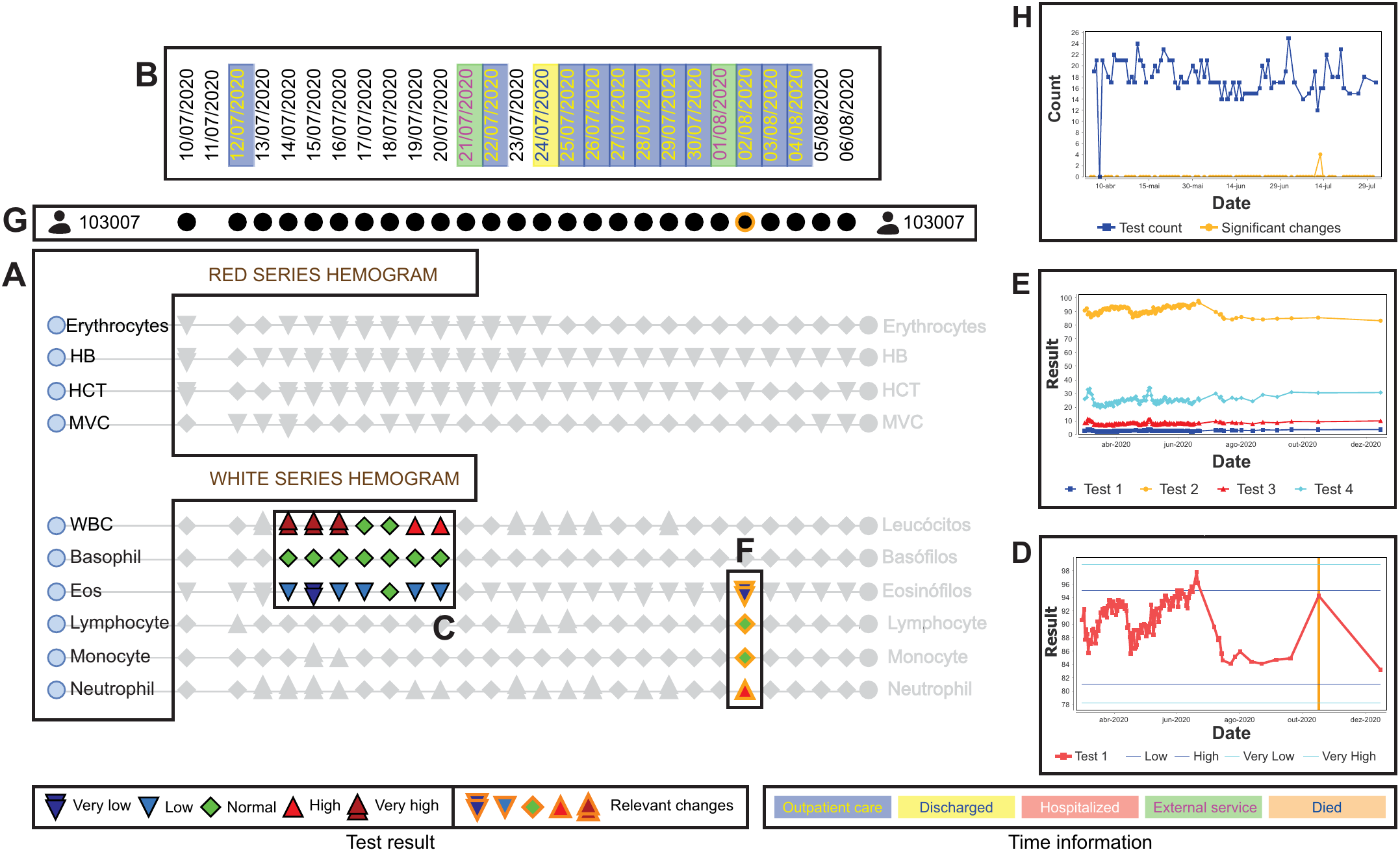}
	\centering
	\caption{Example of the Clinical path visualization divided into the main features: (A) Categorization of the tests; (B) Timestamp information with color codding\,---\,note that the dates follow the Brazilian date format (dd/MM/yyyy); (C) Color and shape codding (symbols) for the test results; and (D) Line chart with the test result variation over time; (E)  Line chart combining different test results over time; (F) Automatic highlighting relevant changes in tests results over time; (G) Patient global and local information over time; (H) Line chart showing patient information over time.}
	\label{fig:clinical_path_didatic}
\end{figure*}

\myParagrapho{Test Results Categorization.} Following the experts' needs, we discretized the test results into five categories, using the reference values as the baseline, as detailed next.
The result is set as normal when it is between the minimum ($ref_{min}$) and the maximum ($ref_{max}$) reference value of this test type. Otherwise, the result is set either to \emph{low} or \emph{high}, and they are split again as \emph{low/very low} or \emph{high/very high}. 
For example, in Fig.~\ref{fig:reference_value} the Mean Corpuscular Volume (MCV) results over $ref_{max}$ were set to ``High'' or ``Very high'' based on the median of values above reference values, evaluated as 98.8. Reference values change individually per patient depending on the patient risk group. The data pre-processing steps that we used were originally implemented in~\cite{icovidvis_cbms} using Python 3.8 and are freely available\footnote{\url{bitbucket.org/gbdi/CovDaSh}}.

\section{\systemname}
\label{sec:clinicalpath}

To evaluate the patient conditions, we proposed the \systemname~system containing several features to improve EHR for physicians' analysis tasks.

\myParagrapho{Visual Design.} Fig.~\ref{fig:clinical_path_didatic} presents the main features of \systemname, which is based on a timeline representation that comprises a high volume of patient information into a simple, intuitive, and compact visualization (T1). We took advantage of several principles used in similar representations, such as ordering, time organization, and interaction with layout, and adapted them to our context. The \systemname~ is visualized separately for each patient, following the structure of single-patient tools. 
The target audience for our tool is composed of physicians, who often deal with diagnostic hypotheses in their daily routine. Examples of diagnostic hypotheses include analyzing requested laboratory tests and hypothesizing which disease is the cause of symptoms in the patient.

Fig.~\ref{fig:clinical_path_didatic}(A) illustrates the groups of tests according to a pre-defined vertical ordering (T6 -- see the supplemental material for details about the groups). The orderings were manually defined by a domain expert, covering all the tests in the data set. The first categories are the most common tests for general physicians and practitioners (\eg, the hemogram), followed by the more specific ones (kidney function, cardio, etc.). The order inside each group may have clinical meaning, approximating tests often analyzed together. However, this ordering may change depending on the context: for example, another order may be more appropriate for a hospital highly specialized in heart disease.

Fig.~\ref{fig:clinical_path_didatic}(B) highlights the clinical history in the time domain according to the clinical path (T5). The timestamp of each patient test is shown in a rectangle whose color is related to the patient outcome in the hospital. It indicates if the patient was hospitalized (red), went to external services (green), outpatient care (blue), was discharged (yellow), or died (orange). It can be helpful to understand, for instance, what was the patient's previous location inside a hospital or possible outcomes, valuable information for guiding a diagnosis. The user can also customize these pre-chosen colors.

\begin{figure*}[ht]
	\includegraphics[width=1\linewidth]{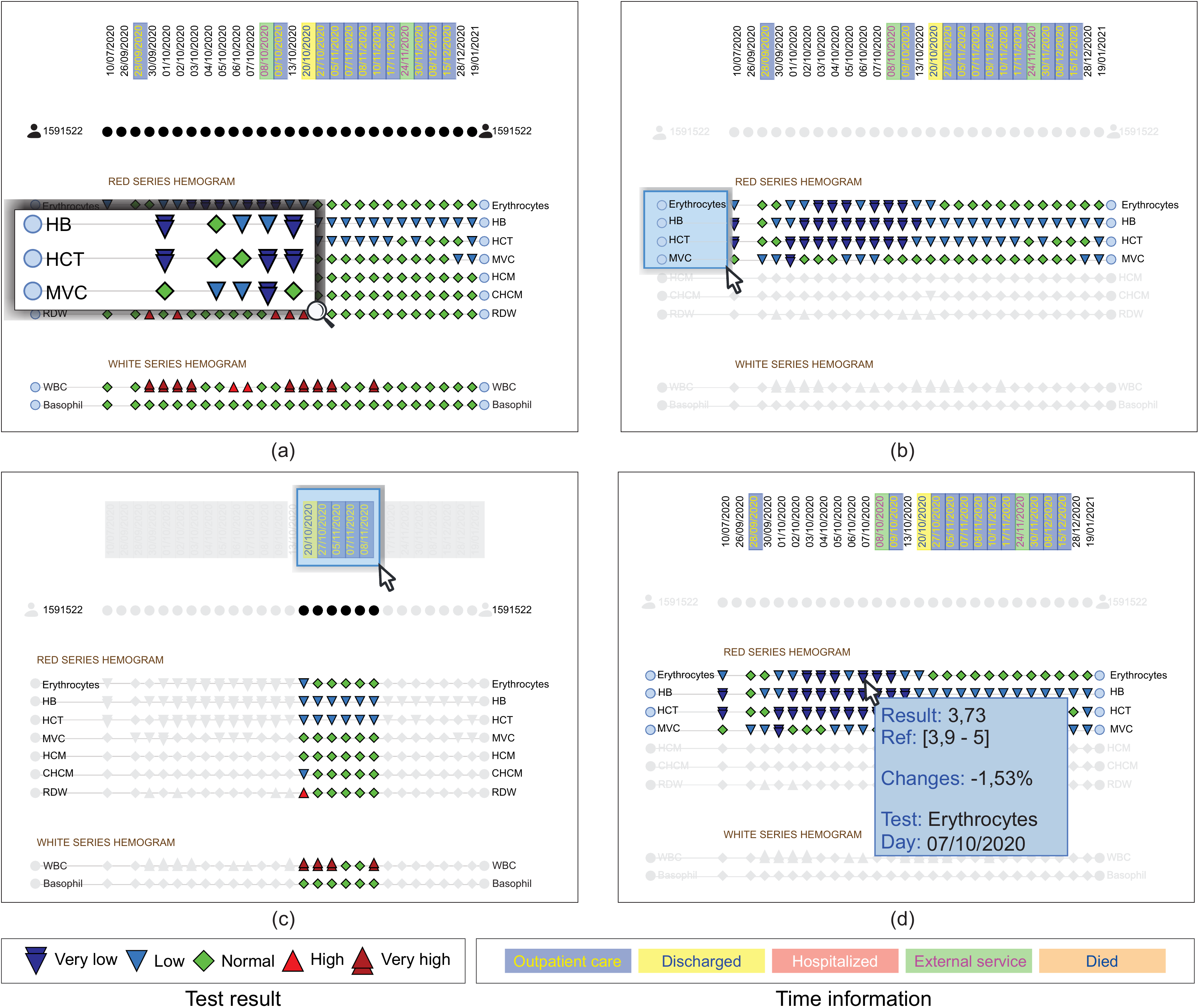}
	\centering
	\caption{User interaction over \systemname: (a) Zoom and pan features; (b) Node selection; (c) Timestamp selection; (d) Tooltip with complementary information. Note that selected and non-selected elements are shown with different opacity levels.}
	\label{fig:user_interaction}
\end{figure*}

Fig.~\ref{fig:clinical_path_didatic}(C) presents the test results in different timestamps according to different colors and shapes (T1, T4). To visualize the test results, we decided to use redundant coding, \ie, to represent the same information in two visual channels (color and shape), in order to avoid color blindness issues and to reinforce the information~\cite{info_visualization}. The test result categorization described in Section~\ref{sec:preprocessing} is visually identified through five symbols: very low (\img{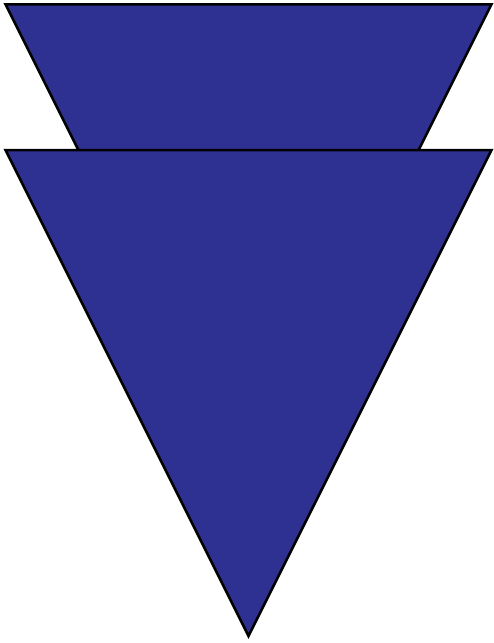}); low (\img{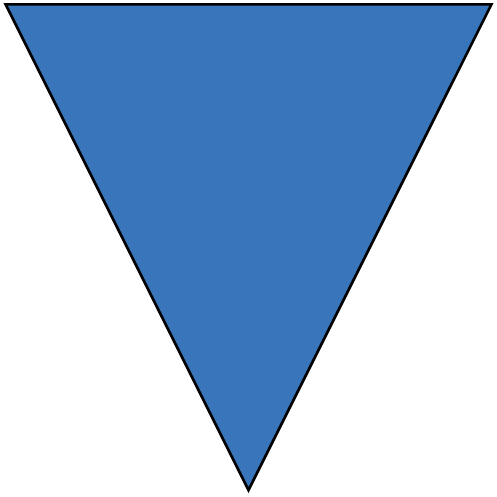}); normal (\img{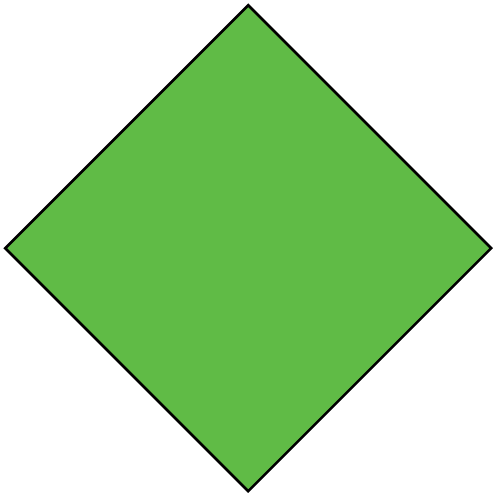}); high (\img{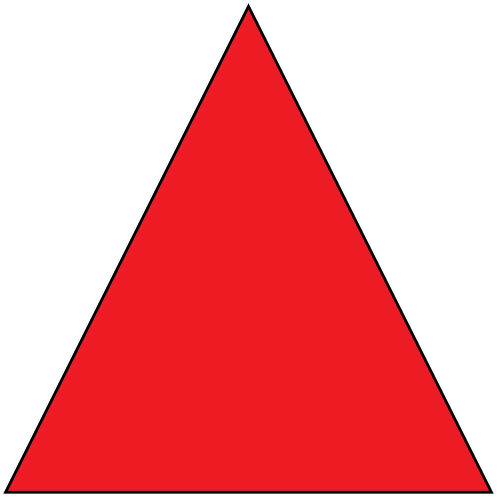}); and very high (\img{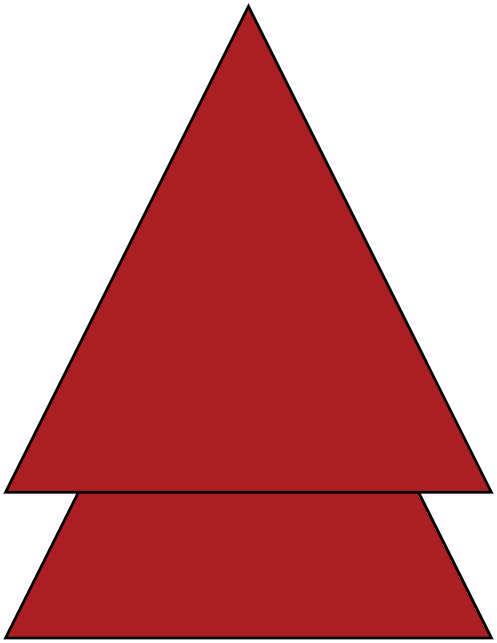}). The colors used for very low/low and high/very high values were chosen as they are commonly used in temperature-based visualizations \,---\, the system also allows color customization according to user preferences. We decided to use this symbol technique as our main visualization since we need to reduce the screen space required without too many interactions, to fit more information (such as textual and numerical values from test results) over time. Although this technique may reduce vertical screen space, it may need more horizontal space, so we decided also to use line charts accessed through interaction as a complementary feature.

Fig.~\ref{fig:clinical_path_didatic}(D) illustrates a line chart, with a red line representing the results of a specific test varying over time (T1, T4). The dark blue lines show the low and high reference values and the light blue lines show the very low and very high values, which help to understand the test results, highlighting whether the results are above or below the defined threshold. We also show orange vertical lines to indicate timestamps with high variation of test results\,---\,this feature will be explained when analyzing Fig.~\ref{fig:clinical_path_didatic}(F).
This visualization is very useful to represent, in small screen space, the entire variation of the results. It enables the selection of regions of interest (timestamps) with interesting variations. This line chart is accessed by double-clicking on the test's name in Fig.~\ref{fig:clinical_path_didatic}(A). Each line chart is opened in a different window and can be manually positioned to compare different tests easily. Alternatively, users can select multiple tests and open them in a single window, as illustrated in Fig.~\ref{fig:clinical_path_didatic}(E). In this case, lines representing reference values and timestamps with high variation are hidden to reduce clutter. Different tests are associated with different colors and symbols along their lines. 

\begin{figure*}[ht]
	\includegraphics[width=1\linewidth]{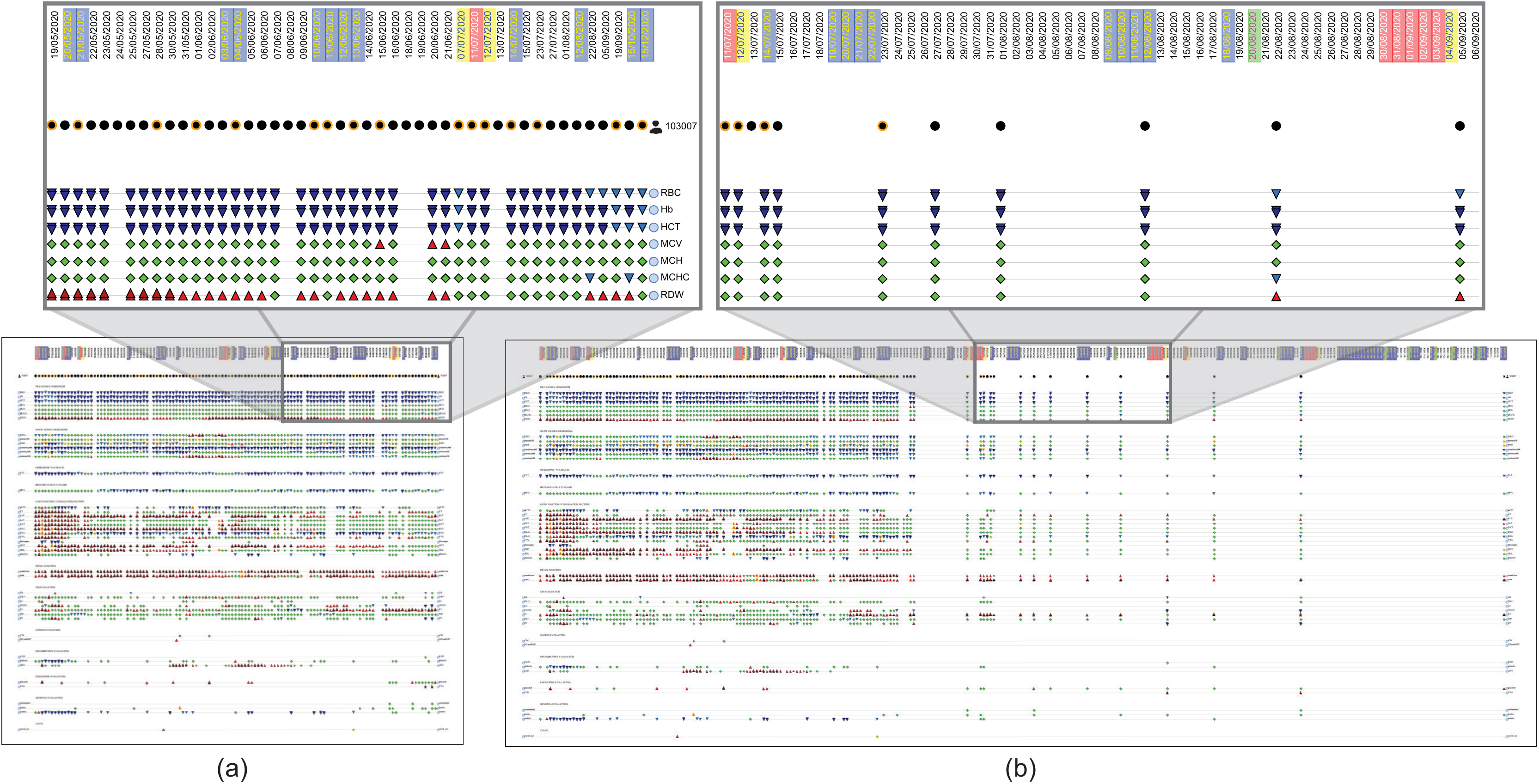}
	\centering
 	\caption{Impact of the filtering option ``Only days with tests'' in the screen space required for a patient with sporadic test results. Filtering enabled (a) and disabled (b). The highlighted areas show the blank space reduction. Note that the zoomed area in (a) comprehends an interval larger than the one from (b) and it uses less screen space.}
	\label{fig:flag_only_days_with_exams}
\end{figure*}

Fig.~\ref{fig:clinical_path_didatic}(F) illustrates a visual approach to automatically highlight relevant changes in test results over time through strong orange borders applied to the symbols defined in Fig.~\ref{fig:clinical_path_didatic}(C). This way, we improve guidance by adding a ``hint'' in the layout that fosters the user to draw his/her attention to meaningful variations in test results, even though such a variation does not imply changes in the result category.
For instance, in the example of the orange vertical line of Fig.~\ref{fig:clinical_path_didatic}(D), the test results drastically increase from the last timestamp but are still in the normal range values. In cases like this, alerting the physician may be important to indicate strong variations and prevent this value from increasing further. This variation was calculated using the rate of change (or percentage change -- $RC$) illustrated in Equation~\ref{rate_of_change}, where $V_{lt}$ is the value at a later time and $V_{et}$ is the value at an earlier time. When there are two values ($V_{et}$ and $V_{lt}$) for two points in time (such as the test results in different timestamps), we can calculate how much the values changed between those two times. To decide if the result of a test in a timestamp is relevant or not, we empirically define a threshold where if the absolute value of the rate of change ($|RC|$) is higher than 100 (which represents an increase or decrease greater or equal to 100\% relative to the last result), the test result had important changes.

\begin{equation} \label{rate_of_change}
RC = \left [ \left ( V_{lt} - V_{et} \right ) - 1 \right ] * 100
\end{equation}

Fig.~\ref{fig:clinical_path_didatic}(G) shows the space dedicated to the patient information, which contains global and local information over time. 
When the person icon \img{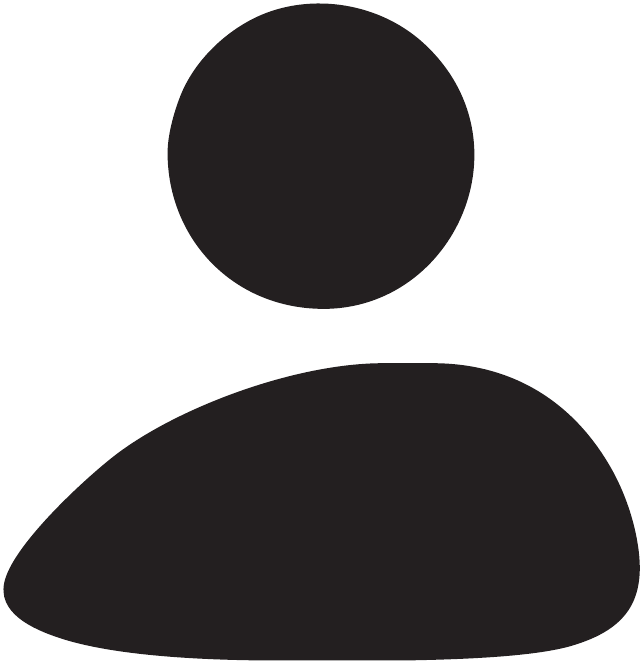} is hovered over, a tooltip with the patient-specific meta-information is displayed. Moreover, the black circles \img{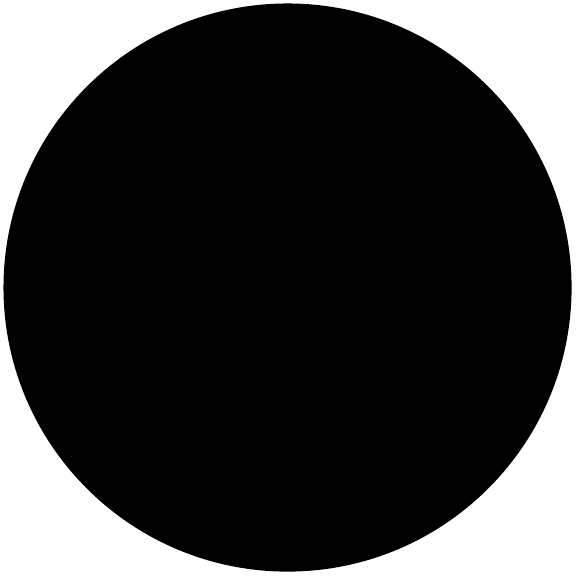} appear in timestamps where there was at least one test and can be highlighted with orange borders \img{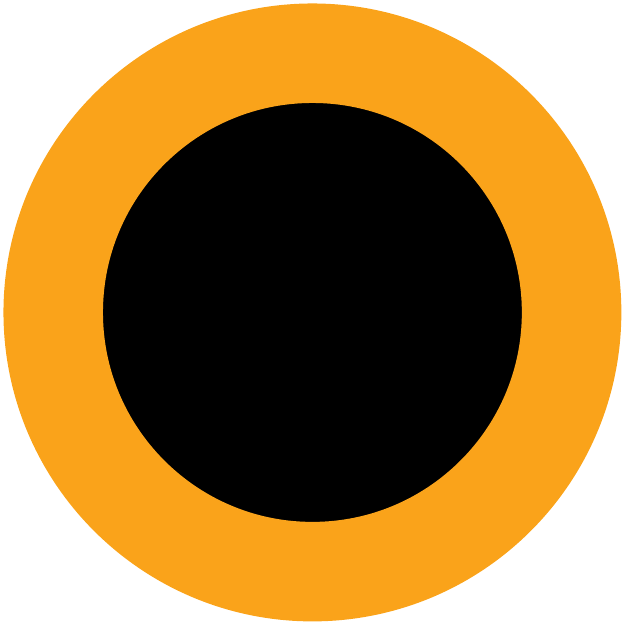} in timestamps where there are tests results with relevant changes. When a circle is hovered over, the system shows a tooltip with a day-to-day summary of the main activities, including the count of tests, the number of normal and abnormal values, and the number of relevant changes at that specific timestamp. Similarly, Fig.~\ref{fig:clinical_path_didatic}(H) presents a line chart summarizing the count of tests and relevant changes over time. All these functionalities assist users in finding relevant timestamps of interest.

\myParagrapho{User Interaction.} We propose several interactive tools to improve the user experience and facilitate navigation and decision-making processes, such as those exemplified in Fig.~\ref{fig:user_interaction}. Among the available interactions, the system allows zooming and panning to facilitate navigation and exploration (Fig.~\ref{fig:user_interaction}(a)). Users can also perform selections by clicking on specific tests/timestamps or drawing a selection area to highlight regions of interest and focus on what is the current target (T2, T4 -- Figs.~\ref{fig:user_interaction}(b, c)). Finally, the system provides extra information about a test result of interest when hovering the mouse over its corresponding symbol, as depicted in Fig.~\ref{fig:user_interaction}(d). 

\begin{figure*}[ht]
	\includegraphics[width=1\linewidth]{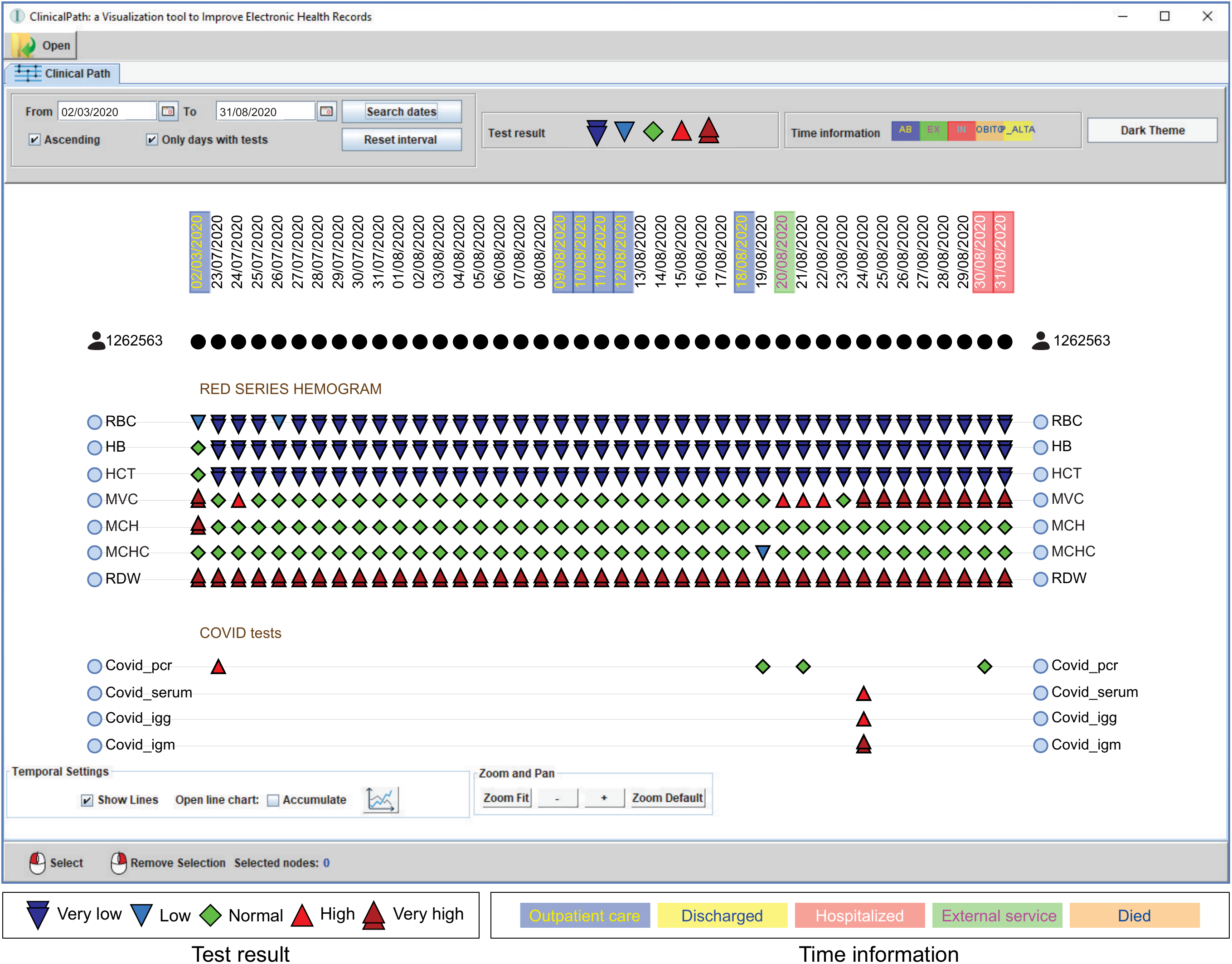}
	\centering
\caption{\systemname~for other patient activity over time, with COVID and red series hemogram test information in the \systemname~system.}
	\label{fig:clinical_path_way}
\end{figure*}

Another set of interactions focuses on date manipulation (T3). Besides the regular options\,---\,such as changing the initial and final date of interest and arranging the days according to an ascending or descending order\,---\,the system offers a filtering option to exhibit only days for which there was at least one test result (flag ``Only days with tests'', see Fig.\ref{fig:clinical_path_way}). This filter is essential to hide sequential days with no activity, thus improving the visualization by horizontally approximating tests' results and reducing the screen space needed. This filter is critical when analyzing patients that performed tests sporadically, as illustrated in Fig.~\ref{fig:flag_only_days_with_exams}. Note that the events in the timeline do not match the actual varying time interval when using this filter. Also, although the hidden timestamps do not have test information, they are part of the patient's clinical history (\ie, they contain time information), which can be misleading. Users are free to choose when this filter should be used. Other actions, such as changing the interface's theme (light/dark), customizing the colors of elements, and exporting results, can be accessed through buttons/flags in the system interface (see Fig.~\ref{fig:clinical_path_way}).

\myParagrapho{Design Decisions.} Many guidelines to improve design have been proposed over the years, for example, the traditional guidance approaches about Gestalt laws, glyphs foundations, Shneiderman's mantra, and others~\cite{10.1111:cgf.13730, 6634158, 7534883, 545307, 10.2312:conf:EG2013:stars:039-063, gestaltvisualization}. We guided our decisions based on those concepts and after considering alternative approaches. As an example, we considered other design choices for showing the test results (Fig.~\ref{fig:clinical_path_didatic}(C)). One would be to show a sequential series of line charts for each test. However, seeing the results variation over time could be difficult depending on the height of this line chart and the number of tests simultaneously displayed. To maximize the number of tests on-screen, we used symbols to represent the tests' results variation over time and opted for allowing access to the line chart through interaction.
Once using symbols to represent tests' results, we had to choose between showing the symbols of all five categories and prioritizing the exhibition of the ``critical'' ones (those representing low/very low or high/very high results). According to our health specialists, since a disease may be characterized by having ``normal'' results for some tests and  ``critical'' ones for others, all five categories should have the same importance to enhance the analysis of diagnostic hypotheses. Our decision, in this case, aims to meet their requirement.

Regarding the timestamp information (Fig.~\ref{fig:clinical_path_didatic}(B)), an alternative approach could be creating different glyphs or symbols to represent each piece of information. Even though some of this time information have intuitive and recognizable glyphs (such as a skull or grave for the ``died'' option), in the case of other values (\eg, outpatient care, discharged, hospitalized, and external service), finding intuitive symbols or glyphs is a non-trivial  task~\cite{10.2312:conf:EG2013:stars:039-063}.
Besides, this information may change from hospital to hospital, making it hard to establish a standard set of glyphs to use. Due to its simplicity and overall acceptance, we thus decided to use colors to indicate time information with the support of tooltips to demonstrate time information values.

At last, regarding the highlighting of relevant changes in tests' results (orange border\,---\,Fig.~\ref{fig:clinical_path_didatic}(F)), other design alternatives could include the use of different symbols' sizes or opacity levels to distinguish between tests with and without relevant changes (binary output). We have also considered using the computed rate of change to overcome this binary output by adjusting the size or opacity level based on its value.
However, besides the discussion about test results symbols having the same importance for analysis, having continuous symbols of different sizes and/or opacity levels would lead to layouts not so pleasant and poorly organized, thus contradicting the proximity and continuity laws of Gestalt~\cite{gestaltvisualization}. In addition, we discarded making changes involving opacity because this property was previously used in several user interactions (see Fig.~\ref{fig:user_interaction}).

\begin{figure*}[ht]
  \centering
  \includegraphics[width=\linewidth]{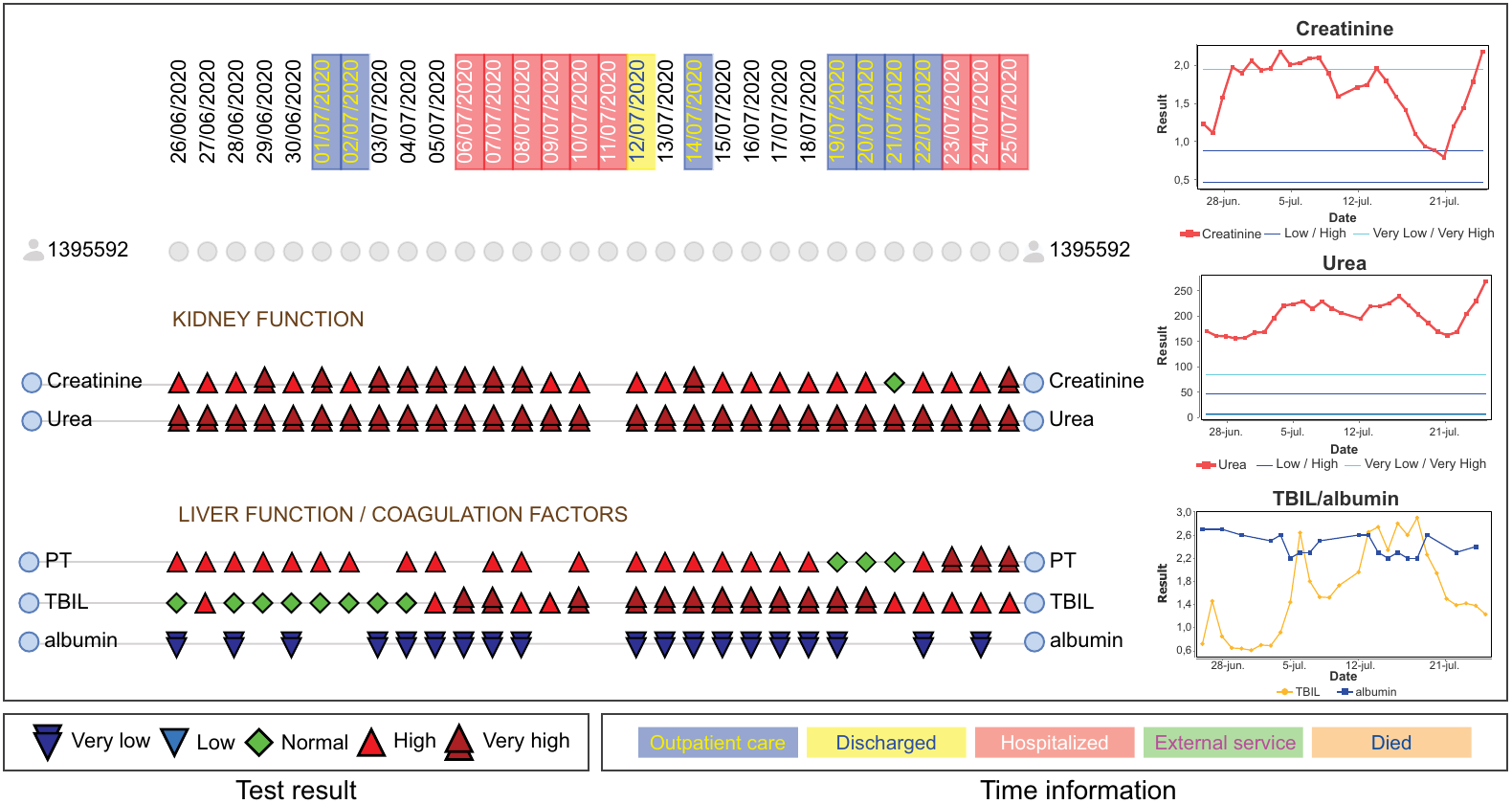}
  \caption{Visualization of patient activity over time, focused on kidney and liver function tests, using the proposed \systemname~system. It shows the tests for creatinine, urea, prothrombin time (PT), total bilirubin (TBIL), and albumin using symbols and a line chart detailing the variation over time. The layout also highlights, at each timestamp, the patient clinical evolution. The results show that this patient had very high values of creatinine and urea tests over several consecutive days, indicating kidney disease. The increase in both TBIL and PT indicates that the liver is not functioning correctly.}
  \label{fig:clinical_path_kidney}
\end{figure*}

\myParagrapho{Implementation Details.} \systemname~is a freely available system\footnote{\url{github.com/claudiodgl/ClinicalPath}} developed in Java to support multiple platforms. 
It also uses the JGraphX library to model the geometric shapes with interaction and JFreeChart to model the interactive line chart. \systemname~also requires the data represented in a graph format, where the source nodes are the patients, the target nodes are tests, and the edges are the test results in each timestamp. Due to the resolution of the Data Sharing/BR repository, each timestamp represents a day. Also, this format helps to understand the relationship between patients and tests. 

\section{Case Study}
\label{sec:case_study}

To choose suitable patient cases for analysis, we worked alongside our domain experts to select cases with enough relevant information to create diagnostic hypotheses and identify essential changes in the tests' results. The cases used throughout this section and the next one were manually selected among those with several performed tests, in which we could create interesting clinical histories.

The \systemname~is a new tool to understand the data evolution and to trace details from the patients' condition in a novel and customizable way. We describe here how the six tasks requested (T1 - T6) and steps of the clinical workflow are fulfilled using the \systemname~in two real scenarios, shown in Figs.~\ref{fig:clinical_path_way} and~\ref{fig:clinical_path_kidney}. 

Fig.~\ref{fig:clinical_path_way} represents a clinical path from a patient's activity over several consecutive days. Two test categories are highlighted: red series hemogram and COVID tests. In the first case, it is possible to see that several tests on different days presented abnormal results. In the case of hemoglobin (HB), the results were very low (\img{/pictures/Inline/MB.pdf}) for several days, possibly leading to an anemic or a reduced total red cell mass state. Also, Red Cell Distribution Width (RDW) presented very high (\img{/pictures/Inline/MA.pdf}) values, suggesting investigation for macrocytic anemia diagnosis, correlated to the red series results. Moreover, due to COVID tests, it is possible to see that the first test of PCR presented abnormal results, indicating that the patient was infected with COVID. Other COVID tests were performed later, with PCR normal (\img{/pictures/Inline/N.pdf}), showing that the patient had recovered from the COVID, but with IgG and IgM abnormal, which indicates that the patient had the disease before, also coherent with the previous PCR test. Also, it is possible to see that the COVID infection may have aggravated the anemia. All this information can be gotten directly from a single screen/chart, allowing the physician to see it all at once.

Fig.~\ref{fig:clinical_path_kidney} illustrates another patient in a critically ill condition. He is 66 years old, hypertensive, diabetic, and has a high body mass index, indicating obesity.
Although depicting complementary measures, there is a correlation between the results shown in the line charts and the symbols. Looking at the timestamp colors, it is possible to see that the patient alternates on some days between hospitalized (red) and outpatient care (blue), indicating the unstable patient condition.
It can be noticed that the creatinine and urea results were tested over several consecutive days, having test results almost always abnormal, especially with urea results very high (\img{/pictures/Inline/MA.pdf}) and creatinine alternating between high (\img{/pictures/Inline/A.pdf}) and very high (\img{/pictures/Inline/MA.pdf}). These results indicate that the patient had critical kidney disease.

Furthermore, concerning the liver function tests, the results of the total bilirubin (TBIL) indicated stability around the first 9 days, alternating between normal (\img{/pictures/Inline/N.pdf}) and high (\img{/pictures/Inline/A.pdf}). However, after 05/07/2020 (date format dd/MM/yyyy), the TBIL results increased drastically, which indicates that the liver was not functioning correctly, leading the physician to decide to hospitalize the patient. After six days of hospitalization, the patient requested hospital discharge due to emotional issues. Since the patient was not in a stable condition, the physician indicated in the plan of care for the patient to maintain regular tests, such as peritoneal dialysis, and make sporadic returns in outpatient care. After the date 21/07/2020, one can notice a high increase in the values of creatinine and urea tests, worsening urinary clearance, high (\img{/pictures/Inline/A.pdf}) values of TBIL, and very low (\img{/pictures/Inline/MB.pdf}) values of albumin, which causes an increase in lower limb edema. Also, with all these previous conditions, the increase of the prothrombin time (PT) on the date 22/07/2020 becomes critical since it suggests a liver failure. In this sense, an important regression of the condition led to a new hospitalization on the date 23/07/2020. In this new case, the physician indicates a therapy in the new plan of care,
such as albumin replacement.

\section{User Evaluation}
\label{sec:user_evaluation}

To validate our \systemname, we performed a user evaluation experiment. We focused on how the system could be used to help a physician perform a diagnosis. We presented two new cases to be evaluated by the participants. They had no previous knowledge about the cases. We focus the evaluation on the main features of the system due to time limitations.

\subsection{Setting the Experiment}

Our study was conducted both online and remotely, with each participant answering the questions at different times and using their personal computers, so we did not have any control over the environment of the study. We provided each participant with two videos. One is a tutorial about the system installation steps, and the other contains instructions about the system's basic functionalities. The functionalities video was 5 min long, and continuously available during the experiment.  

\subsection{Online Questionnaire}

All questions were created or validated by a domain expert physician to avoid misleading and to provide the correct acronyms, terms, and knowledge. The questionnaire was written in Brazilian Portuguese, which is the native language of all participants. Every test's names, categories, diagnostics, and descriptions were converted to the participants' native language. The questions were divided into (i) background and experience; (ii) three basic questions, \ie, questions whose answers do not require medical knowledge from the participants; (iii) five advanced questions, \ie, questions that require medical knowledge to be answered, containing a fictional patient's medical history (four multiple-choice and one open question); (iv) Likert-scale-based questions to evaluate the user preference of the system, along with an open question to justify the choices; and (v) a mix of multiple choices and open questions to collect the users' feedback about the system. A complete description of the questions and possible answers is available in the supplemental material. This structure for the questions was based on similar user studies evaluating layouts or systems in the visualization field~\cite{8440810, 9716779, 8440832, Linhares2021}.

We formulated the simple questions to validate if the participants understood the basic functionalities of the tool and the visual encoding proposed in this work. The goal behind the advanced questions was to validate if the participant could analyze a patient's condition based on the fictional description with a clinical history and the available test results. This clinical history contains extra information about the patient's gender, age, symptoms, and current condition. We refer to the advanced questions by the acronyms AQ1, AQ2, AQ3, AQ4, and AQ5. The objective of the four first questions (AQ1-AQ4) was to guide the participant in specific regions of interest in the layout. These four questions asked the user to validate whether they agree with the described diagnostic hypothesis and the relevant changes in the tests. Finally, we required the participants to fill an open question (AQ5) to list if they found any additional diagnostic hypotheses and/or relevant changes. The open question was intended to encourage users to explore the system and freely look for more patterns.

\subsection{The Pilot Study}

The first experiment was conducted with two participants (not included in the final analysis) as a pilot evaluation to obtain an initial assessment of the response time, the general system correctness, the questionnaire understanding, the instructional video adequacy, and the difficulty level of the advanced questions. From the feedback received, we realized the need for the installation video. In the first version of the experiment, we had five advanced multiple-choice questions. After that, we removed a question classified as hard by the two participants, replacing it with an open question that stimulates the participant to perform an exploratory analysis of the layout. The participants considered the response time satisfactory and the instructional video enough to understand the system's usability.

\subsection{Participants}

We recruited 15 participants, all medical graduates in Brazil. Among the participants, 66\% are men and 33\%women. 60\% of the participants are 24 years old or less. We also asked whether they had previous experience working in health care outside the medical residency, with most of them giving positive answers. 
They participated voluntarily in the experiment.

\subsection{Results}

We carefully prepared the questions to require only basic medical knowledge from the participants to evaluate the layout quality. Therefore, the advanced questions were designed not to require previous expertise and lead to a more straightforward answer. The first four advanced questions have three answer options: ``yes'', ``no'', and ``I don't know''. They were designed so that when the participant chose the ``I don't know'' option, it is safe to assume that they did not remember what was required to evaluate that point, so we do not compute this question. To exclude the possibility that the ``I don't know'' answer means that some part of the layout was impairing the analysis, we asked them to provide feedback in an open question to evaluate the real reason. Of the 60 available answers from the advanced multiple-choice questions (four questions from every participant), six answers were marked as ``I don't know''. Hence, we assume that the participants had the previous knowledge to evaluate the study in most cases. Also, the participants spent, on average, 28 minutes answering all the questions, with a standard deviation of 24 minutes and 14 seconds, indicating a significant variance between the response times. 

We evaluated the analysis following two aspects of the participants, according to their self-description attributes. Then, we divided each group into subgroups with different characteristics (Table~\ref{tab:groups}). The group sizes were always relatively similar. Interestingly, the total time to execute the experiment was similar to the average time for all participants (27min36s), with a high variation, except for the female subgroup. The fastest subgroups to complete the experiment were the older ones and females. 

\begin{table}[ht]
\centering
\caption{Groups with attributes, size, and minutes to complete the questionnaire. The symbol $\pm$ represents the standard deviation.
}
\label{tab:groups}
\resizebox{0.4\textwidth}{!}{%
\begin{tabular}{llll}
\hline
\textbf{Groups}                          & \textbf{Subgroup}          & \textbf{Size} & \textbf{Total time}     \\ \hline
All participants              & -- & 15    & 27.36 $\pm$ 24.14 \\ \hline

Age group              & $\leq$ 24 & 9    & 30.26 $\pm$ 27.11 \\
                                & $\geq$ 25 & 6    & 23.19 $\pm$ 20.37 \\ \hline
Gender                 & Male              & 10   & 28.36 $\pm$ 29.42 \\
                                & Female            & 5    & 25.36 $\pm$ 7.57  \\ \hline
\end{tabular}%
}
\end{table}

Table~\ref{tab:correct_answers} presents the proportion of correct answers for the multiple-choice questions. The results were divided into simple and advanced questions. The accuracy of the simple questions was 100\%, \ie, all participants provided the correct answer for the three questions based on the visualization, which did not require previous medical knowledge. For the advanced questions that required medical knowledge, the participants achieved a success rate of almost 83\% considering all participants. AQ2 presented the highest and AQ3 presented the lowest accuracy among the four questions. We had no control over the environment of the study to acknowledge why participants had more difficulty in AQ3 than in other questions. We hypothesize that the display resolution and the need for screen scrolling to analyze multiple tests (recall that this question demanded more comparisons of tests than the others) may have negatively impacted AQ3 results. Considering the simple and advanced questions, the answers of all participants achieved a satisfactory value of 90,4\%. The participants that achieved the best results were females and those less than 24 years old. The female subgroup achieved 94,2\% of correct answers with no question marked as ``I don't know''. In addition, the older group had the worst result of correct answers and the highest percentage of ``I don't know'' choices.

\begin{table}[ht]
\centering
\caption{Percentage of correct answers in the questionnaire according to simple and advanced categorization and also divided into the two groups. IDK represents the questions' answers as ``I don't know''. First-line indicates the answer of all participants. The following ones are the two group divisions from Table~\ref{tab:groups}.}
\label{tab:correct_answers}
\resizebox{0.49\textwidth}{!}{%
\begin{tabular}{lcccccccc}
\hline
\textbf{Correct (\%)} & \textbf{Sim.} & \textbf{Adv.} & \textbf{AQ1} & \textbf{AQ2} & \textbf{AQ3} & \textbf{AQ4} & \textbf{IDK} & \textbf{Total} \\ \hline
All participants        & 100           & 83,3          & 86,6         & 93,3         & 73,3         & 80           & 10           & 90,4           \\ \hline
$\leq$ 24     & 100           & 94,4          & 88,8         & 100          & 100          & 88,8         & 5,5          & 96,4           \\
$\geq$ 25     & 100           & 66,6          & 83,3         & 83,3         & 33,3         & 66,6         & 16,6         & 80,9           \\ \hline
Male         & 100           & 80            & 90           & 90           & 60           & 80           & 15           & 88,5           \\
Female       & 100           & 90            & 80           & 100          & 100          & 80           & 0            & 94,2           \\ \hline
\end{tabular}%
}
\end{table}

In the advanced questions using the patient's clinical history, we also asked the participants, in an open question, if they found any additional diagnostic hypothesis and/or relevant variations in the test results. Among them, 40\% mentioned at least one new result. Four participants cited the elevation of Blood Glucose indicating Diabetes Mellitus, and two participants cited the Hyperkalemic Acute Renal failure and Respiratory Alkalosis. Moreover, other possible problems raised were Hyperkalemic, Thrombophilia, Hereditary Coagulation disease, and Placental bed subinvolution. These results found in the exploratory analysis were validated by a physician according to the clinical history and the test results and were classified as correct answers.

\begin{figure}[ht]
	\includegraphics[width=1\linewidth]{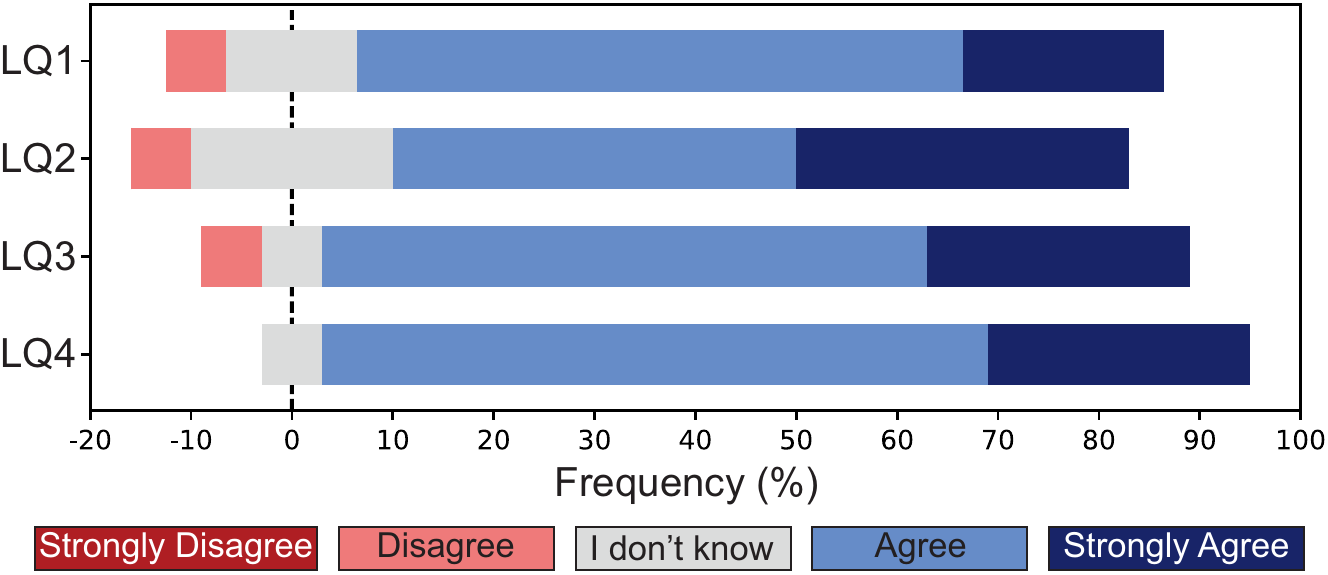}
	\centering
	\caption{Participants answer for the user preferences of the system. The bar length is the percentage of respondents who chose a specific Likert level. The questions descriptions are LQ1 -- The interface and usability of \systemname~is intuitive; LQ2 -- \systemname~is efficient and would optimize my time; LQ3 -- \systemname~is easy to use; LQ4 -- It's easy to learn how to use \systemname.}
	\label{fig:likert}
\end{figure}

We also asked about the participants' preferences about the system using a 5-point Likert-scale questionnaire (Fig.~\ref{fig:likert}). We divided the choices into four questions requesting the level of agreement concerning the interface intuitiveness and usability (LQ1), the efficiency of the system to optimize their analysis time (LQ2), ease to use (LQ3), and system learning easiness (LQ4). Figure~\ref{fig:likert} uses diverging stacked bar charts, which is recommended for this data type~\cite{Linhares2021}. Blue bars indicate the participants' agreement and red ones the disagreement. The bars are centralized at the neutral options (gray color), which is the ``I don't know'' option. Then, we converted the results to the frequency (percentage) of participants that agreed or disagreed with the four raised questions (LQ1 - LQ4). In general, related to the intuitiveness of the \systemname, we achieved more than 80\% of positive answers (agree or strongly agree) for the LQ1 question, validating the intuitiveness of our system. Three distinct participants diverged from the others, differing in the first three questions, each one for each question. All other participants only marked neutral or agreement. Finally, among the 60 available answers, 50 were marked as ``Agree'' or ``Strongly Agree'', resulting in a satisfactory level of 83\% positive responses. 

To understand the user preferences on the 5-point Likert scale, we also collected an open question asking them to justify their choices. First, we analyzed the answers of the three participants that marked ``disagree''. The participant with the negative choice about the usability (LQ1) suggested some improvements claiming that \textit{``to keep the test and time labels during scrolling would help the physician''}. The participant that disagrees with the question about the efficiency (LQ2) suggested integrating the system with a textual medical record system that he/she was familiarized with. At last, the participant that disagreed with the easiness of using the system (LQ3) justified his/her choice, saying that \textit{``if I have had more familiarization time, it should become easier to use''}. Another concern mentioned that the dates were not easy to visualize. These two aspects regarding the tool improvements will be integrated into future versions of the system. 

Regarding the positive aspects mentioned, some stated that the system could save time, such as \textit{``Great tool, very practical. It would greatly save time for the patient's tests analysis''}. Some other participants also commented about the symbols and the temporal evolution: \textit{``The idea of graphically demonstrating the reference values through symbols makes it easier to learn the patient's clinical condition over time''} and \textit{``It's a good system, especially because it uses different colors and patterns that allow quick visualization for decision making''}. Other participants mentioned that the tool would fit for clinical use, claiming that \textit{``the tool is intuitive, easy to use, and with great potential for clinical application. I recommend its use''}. 

We also collected the participants' opinions about the most useful visual aid offered by \systemname. Although some answers overlap with the motivations for the positive points in the 5-point Likert scale, the participants were more specific about their visual aspects preferences. Among the 15 responses, 12 mentioned the colors and symbols (arrows ups and down) to represent the test results, which was the most preferred visual resource provided by the system. Others mentioned the longitudinal aspect (the variation over time), the subtitles, the patient data summary on the screen, the graphics, and the test's selection.

\subsection{Other Support Systems Familiar to Participants}

Traditional systems that our domain experts (authors and participants of our experiment) use to explore diagnostic hypotheses in their daily activities are connected with private systems, so they were not available to create direct comparisons. However, we asked some professionals to describe these systems in detail to understand how they work and differ from our proposal. First, from a visualization point of view, they only show information in text and tabbed form. Second, they are limited to a list of tests by a single date. To compare different dates, users need to open new tabs (one for each day or requested test).

When the changes in tests' results are significant, and there is a need to compare many tests and dates, the time required to go through each test (which can represent a tab in the traditional system) makes the task more difficult and time-consuming. In addition, this system of tabs makes it challenging to see the whole picture. Moreover, to perform the analysis of medical records, physicians manually insert the results of altered tests into the system. Since they do not have a tool for comparing results on a single screen, they have to spend even more time with these tasks.

Furthermore, it is difficult to compare patient history in these traditional EHR systems since the medical record is closed at the end of each hospitalization or service for billing purposes. Overall, these systems are focused on meeting the requirements of some hospital sectors (\eg, administrative and financial departments) and not on the requirements of physicians (for the most part), unlike \systemname. Ideally, our \systemname~system can be integrated into those EHR systems to get the best of both worlds.

\section{Discussion and Limitations}
\label{sec:discussionlimitations}

As our case study and user evaluation demonstrated, \systemname~is an intuitive and important tool for physicians to perform decision-making tasks. However, there are limitations in both the data set and the system that require further improvements.

\myParagrapho{Databases.} The Data Sharing/BR repository lacks a detailed description of the variables, laboratory methods involved, acquisition and calibration protocols, and measurement errors. This is aggravated by simplified internal vocabularies, varying notations, and optional observations written in free text, which impairs treating every data at once in a consistent way. Because of this, we needed to filter out several tests that could help identify a hypothetical diagnosis or determine the patient's condition. However, this is the typical scenario of real health databases. Moreover, since the data are anonymous, they lack previous patient history, leading us to create a fictional one for the questionnaire evaluation, which could not perfectly represent the actual patient condition. On the other hand, eventual missing tests or attributes (due to the substantial reduction in the cleaning and normalization steps) had little or no impact on our final analysis because we guided the participants to find relevant cases that focused on the available tests for hypothetical diagnosis in our fictional patient history, reducing the importance of unavailable data. This should not be a problem in real-world scenarios either, since physicians of a hospital would be dealing with the cured data from patients in that hospital and the necessary tests can be ordered. Also, due to the lack of patient information about prescriptions, drugs in use, and follow-up outcomes, we do not include a special visualization metaphor for these information items, which is a topic for future development.

Since new test results are generated every day in real-world scenarios, a relevant aspect is related to incorporating them into the system. Considering that the Hospital Information System (HIS) integrates the hospital's databases, \systemname~would naturally exhibit these new results once they are inserted into the database being used. Also, \systemname~could be a new module to fit into the HIS backbone. Even though \systemname~current version accesses the database in a ``read-only'' fashion, additional extensions could change this direction, so physicians could also make annotations using the system and feed the databases. The integration with the HIS is a task we are very interested in and which would require further technical efforts and domain decisions from both sides.

\myParagrapho{Image Data.} Integrating test images in the system (\eg, Resonance, X-ray, and Tomography tests) would greatly help in validating diagnostic hypotheses and comparing the development of abnormal findings (\eg, damage control of aspiration pneumonia). However, EHRs usually contain primarily information relevant for billing purposes that do not reflect the patient's well-being, whether the patient has agreed to specific treatments or not, and many other aspects that are relevant for treatment decisions, such as image data~\cite{7111928, floricel2021thalis}. Because of that, this case was out of the scope of this work. As a future direction, our system can be adapted to consider these images, adding them as circular points in the layout at specific timestamps and sorting them into respective test categories. These circles could lead to a new window where user interaction demonstrates the image in full resolution. These functionalities need further testing and user validation to confirm their usability.

\myParagrapho{Visual Scalability.} Although the visual scalability was not directly certified in this study, one of the patients that we employed in the user evaluation had 46 tests and 448 timestamps (more than a year with daily information), containing almost 10,000 tests distributed on different days and resulting in a reasonable amount of information. This volume of data is expected since the tool is helpful in following patients in critically ill conditions. Despite this, participants in the user evaluation did not complain about the amount of information on the screen. On the contrary, they positively highlighted the volume of simultaneous details that the system can support. Moreover, we provide several filters for the timestamps, such as searching for specific dates and showing only dates with tests as there are only 73 different tests after the cleaning process (see Section~\ref{sec:preprocessing}), and only a few others may be included. We do not implement filtering by test, considering the categorization of tests as sufficient for the analysis. Furthermore, such categorization is well-suited for the analyzed clinical paths and general practitioner tasks. However, new categories may be required in different contexts, such as physicians from other specialties.

\myParagrapho{Visual Improvements.} According to the user evaluation, there are minor aspects of the visualization that can be improved. In our questionnaire, we specifically asked the participants to suggest improvements. Four participants suggested changing the date representation by moving the labels to the horizontal instead of the vertical position to facilitate the visualization. However, the vertical representation saves screen space and enables the longitudinal comparison, which leads to a conflicting decision. Another possible solution is to change the inclination of the dates in a 60 or 70-degree tilted fashion, to improve the readability and reduce the vertical space needed.
Moreover, the test result symbols (arrows up and down) are designed to quantitative values with a reference range. For tests with qualitative measurements with categorical results, this range may not exist and the test result design needs to be adapted. A possible solution is to use circular points to indicate timestamps where there is a test (similar to the suggestion for image data) and show its corresponding category in a tooltip through user interaction. Also, as we converted the data set to a temporal resolution of a single day (\ie, each column comprehends a one-day interval), there were no two or more test results at the same time. In a case where it is important to have this information in a different temporal resolution (\eg, seconds, minutes, or hours), such as when dealing with real-time information of patients in an intensive care unit, the visualization would naturally meet the new time scale once the data set is converted to the desired resolution. Also, we empirically defined a threshold of 100\% for the rating of change. However, other values can also be interesting for the users to analyze. We intend to keep this default threshold in the system for future improvements and allow users to adjust this value dynamically.

Other changes can improve the layout. In the current version, when a scroll bar is required during zoom interaction, the label information (name of test and date) disappears depending on the region of interest. Although we have tried to overcome this issue by placing this information inside the tooltip, some users found it more intuitive always to maintain the label information visible, which we intend to implement in a future version. Also, participants suggested changing shortcuts: in the current version, the zoom in/out is controlled with the mouse wheel; however, some participants said that this shortcut should be associated with panning instead of zoom, more likely similar to the systems they had used before. Furthermore, we intend to adapt our tool as a web application to make it easier to run in different environments and remove the installation step. In addition, we plan to keep working with domain experts to improve our system data gathering and visualization resources.
Finally, thanks to the reviewers' comments and a meeting with specialists, we included some new features into the system after the user evaluation (relevant changes and the combination of different tests in a single-view line chart). Although these new features were not part of the user evaluation and did not directly affect the reported results, informal talks with physicians highlighted the relevance of these functionalities, which we plan to test in a follow-up study. 

\myParagrapho{Adaptation to Population-based Tool.} Another future extension of the \systemname~is incorporating multiple patient histories in the system. One possible solution would be to select one or more tests to compare several patients for specific tests instead of selecting patients in the initial screen. Instead of comparing multiple tests of a single patient, we would compare one or more tests for all patients. The \systemname~visualization presents a list of tests grouped by different test categorizations in our current implementation. One possibility to adapt our visualization to a population-based tool would be to group patients according to the performed tests. That would enable, for example, the comparison of patients that performed the same tests (probably with similar illnesses). The same tests could be selected and exhibited in the same line chart, similar to Fig.~\ref{fig:clinical_path_didatic}(E). In this way, the system would require a new set of interaction techniques, which would require a new user evaluation to provide proper validation.

\section{Conclusion}
\label{conclusion}

Information visualization plays a crucial role in the understanding and interpretation of data. Moreover, Electronic Health Records (EHR) systems are also essential to facilitate a physician's perception of the patient's condition and help diagnose tasks. EHR combined with visualization techniques can be the basis of insightful and informative systems. In this paper, we proposed and evaluated \systemname, an important tool for EHR that allows users to track the clinical evolution of a patient, revealing his/her test information over time in just a screen view. Our proposal was developed in close collaboration with domain experts to ensure that the technical aspects effectively match the real needs of medical practitioners. In our case study, we analyzed two patients with potential kidney and COVID/anemia issues.
Moreover, we evaluated our proposal with a group of 15 domain experts. We validated our visualization system using multiple-choice questions to verify whether participants could use the tool and find the correct answers for the proposed tasks. We also asked open questions to confirm whether participants could generate new insights in exploratory studies. In total, the participants achieved a performance of more than 90\% correct answers, with 83\% of participants considering only positive aspects, including that the system is easy of learning, with a high level of usability, efficiency, and intuitiveness.

% \balance
% \clearpage

 \acknowledgments{
This research was supported by grants \#2020/10049-0, \#2020/07200-9 and \#2016/17078-0 from Sao Paulo Research Foundation (FAPESP); grant \#312483/2018-0 from the Conselho Nacional de Desenvolvimento Cientifico e Tecnologico (CNPq) and Coordena\c{c}\~ao de Aperfeicoamento de Pessoal de Nivel Superior (CAPES); grant \#E-26/201.424/2021 from Rio de Janeiro Research Foundation (FAPERJ); project AIMED-CATI 030/2007 FOXCONN-001/2019, Foxconn Technology Group, Zerbini Foundation and Getulio Vargas Foundation.
}

\bibliographystyle{abbrv-doi}

\bibliography{paper}
%\balance
 
\vfill\eject

\vbox{%
\begin{wrapfigure}{l}{70pt}
{
%\vspace*{-50pt}
	\includegraphics[width=1in,height=1.25in,clip]{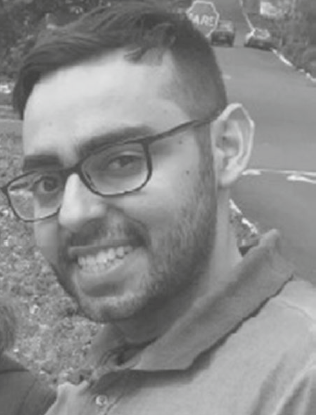}
	%\vspace*{15pt}
}%
\end{wrapfigure}

\noindent\small 
\\
\textbf{Claudio D. G. Linhares}
is a postdoctoral fellow in ICMC at the University of Sao Paulo, Brazil. He received the BSc (2012), MSc (2016), and Ph.D. (2020) degrees in computer science from the Federal University of Uberlandia, Brazil. His research interests include human-computer interaction, information visualization, network visualization, and visual scalability, with applications especially in healthcare.}

\vbox{%
\begin{wrapfigure}{l}{70pt}
{
\vspace*{50pt}
	\includegraphics[width=1in,height=1.25in,clip]{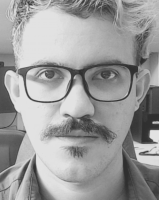}
%	\vspace*{15pt}
}%
\end{wrapfigure}
\noindent\small 
\\
\\
\\
\\
\\
\\
\\
\textbf{Daniel~M.~Lima} is a PhD student at the ICMC, University of São Paulo, Brazil. Graduated in information systems (2008) at the Piaui Federal Institute, and received the MSc in computer science (2013) from ICMC/USP. His research interests include databases, information visualization, and deep learning, especially applications in automation, energy, and healthcare.}

\vbox{%
\begin{wrapfigure}{l}{70pt}
{
\vspace*{50pt}
	\includegraphics[width=1in,height=1.25in,clip]{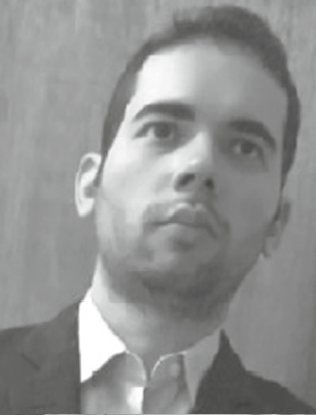}
%	\vspace*{15pt}
}%
\end{wrapfigure}
\noindent\small 
\\
\\
\\
\\
\\
\\
\\
\textbf{Jean R. Ponciano} is currently a postdoctoral fellow at the School of Applied Mathematics of Fundação Getulio Vargas, Brazil. He received
the BSc (2012), MSc (2016), and Ph.D. (2020) degrees in computer science from the Federal University of Uberlandia, Brazil. His research interests include information visualization, visual analytics, network science, and data streams. He has served as a program committee member or external reviewer in relevant conferences, including IEEE Vis, EuroVis, and ASONAM.}

\vbox{%
\begin{wrapfigure}{l}{70pt}
{
\vspace*{30pt}
	\includegraphics[width=1in,height=1.25in,clip]{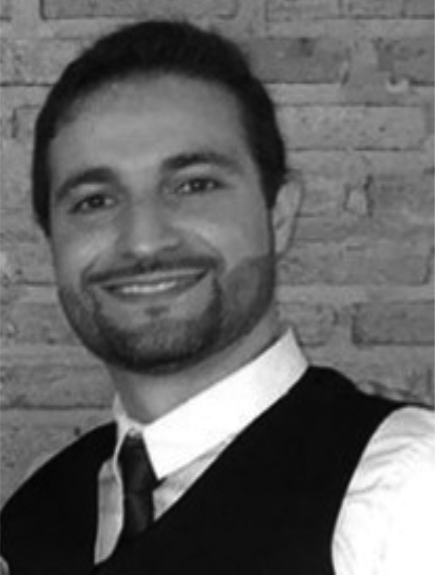}
%	\vspace*{15pt}
}%
\end{wrapfigure}
\noindent\small 
\\
\\
\\
\\
\\
\textbf{Mauro~M.~Olivatto}
is a Medical Doctor student at the Federal University of Fronteira Sul (2016-2022) and a nursing graduated, with hospital experience (since 2006). He has worked as a lecturer of the technical nursing course (Faculdade SENAC-Chapecó, since 2018), works as a first aid instructor at the same institution, and a lecturer of the technical course in Radiology (Escola Técnica Danoliper-2018). In addition, he worked with medical record systems from different institutions - Hospital Regional do Oeste, Hospital Regional São Paulo, Santa Casa de Misericórdia de São Carlos, USF (Family Health Unit) in the following cities: Chapecó-SC, Xanxerê-SC, Marema-SC and Pinhalzinho-SC.}

\vbox{%
\begin{wrapfigure}{l}{70pt}
{
\vspace*{13pt}
	\includegraphics[width=1in,height=1.25in,clip]{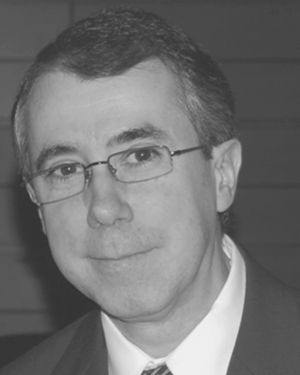}
%	\vspace*{15pt}
}%
\end{wrapfigure}
\noindent\small 
\\
\\
\\
\textbf{Marco~A.~Gutierrez}
received the B.E.E and Ph.D. in electrical engineering both from the University of Sao Paulo, Sao Paulo, Brazil in 1985 and 1995, respectively. He is currently Director of the Biomedical Informatics Laboratory, with the Heart Institute (InCor), University of Sao Paulo and Fellow of the International Academy of Health Sciences Informatics. His research interests include biomedical signals, image processing and visualization.}

\vbox{%
\begin{wrapfigure}{l}{70pt}
{
\vspace*{35pt}
	\includegraphics[width=1in,height=1.25in,clip]{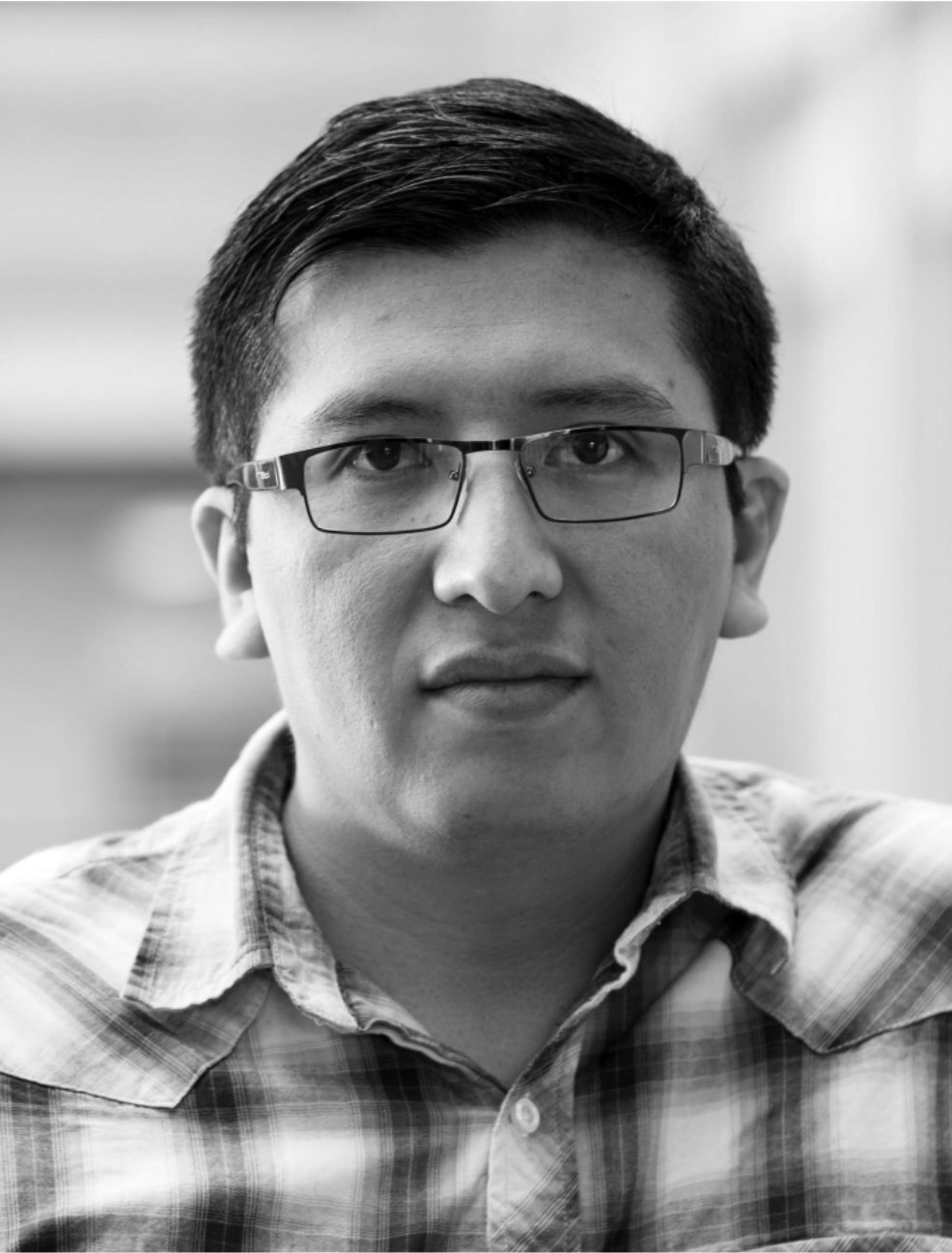}
%	\vspace*{15pt}
}%
\end{wrapfigure}
\noindent\small 
\\
\\
\\
\\
\\
\textbf{Jorge~Poco} 
is an Associate Professor at the School of Applied Mathematics at Fundação Getúlio Vargas Rio de Janeiro-Brazil. He received his Ph.D. in Computer Science in 2015 from New York University, his M.Sc. in Computer Science in 2010 from the University of São Paulo (Brazil), and his B.E. in System Engineering in 2008 from the National University of San Agustín (Peru). 
His research interests are data visualization, visual analytics, machine learning, and data science. He has served on several program committees, including IEEE SciVis, IEEE InfoVis, VAST, and EuroVis.}

\vfill\eject

\vbox{%
\begin{wrapfigure}{l}{70pt}
{
%\vspace*{13pt}
	\includegraphics[width=1in,height=1.25in,clip]{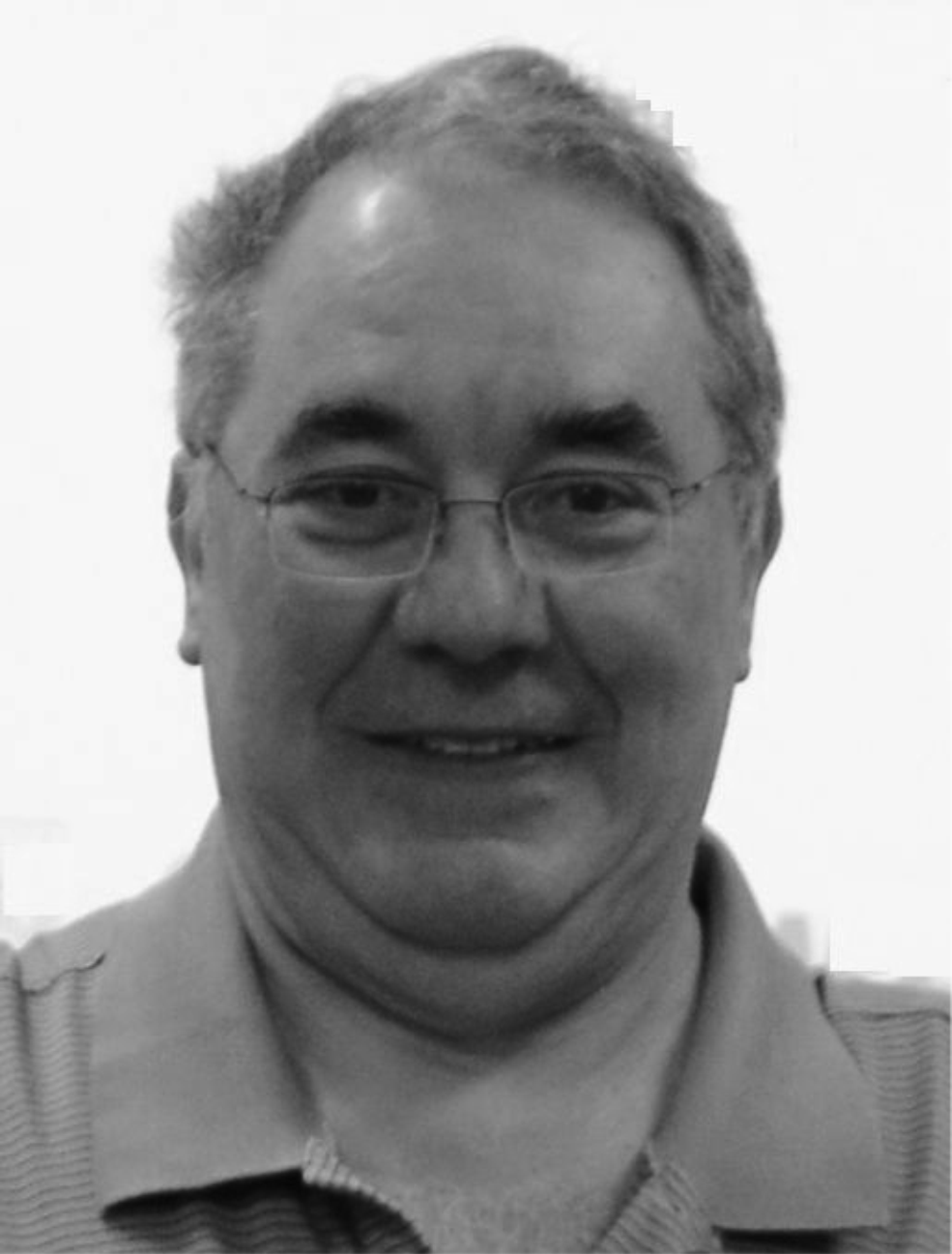}
%	\vspace*{15pt}
}%
\end{wrapfigure}
\noindent\small 
\\
\textbf{Caetano~Traina~Jr.} 
received his PhD in Computational Physics from the University of São Paulo at São Carlos, Brazil, and got his MSc in Computer Science and his BSc in Electrical Engineering from the University of São Paulo at São Carlos, Brazil, in 1986, 1982, and 1977, respectively. He is a Full Professor with the Computer Science Department at University of Sao Paulo at São Carlos.
His research interests include similarity queries on complex data, multimedia databases, data science, and big data querying and analysis.
He has already supervised over 70 graduate students in these areas, and published more than 250 papers in journals and conferences. He is a member of the Brazilian Computer Society, ACM and IEEE Computer Society.}

\vbox{%
\begin{wrapfigure}{l}{70pt}
{
\vspace*{13pt}
	\includegraphics[width=1in,height=1.25in,clip]{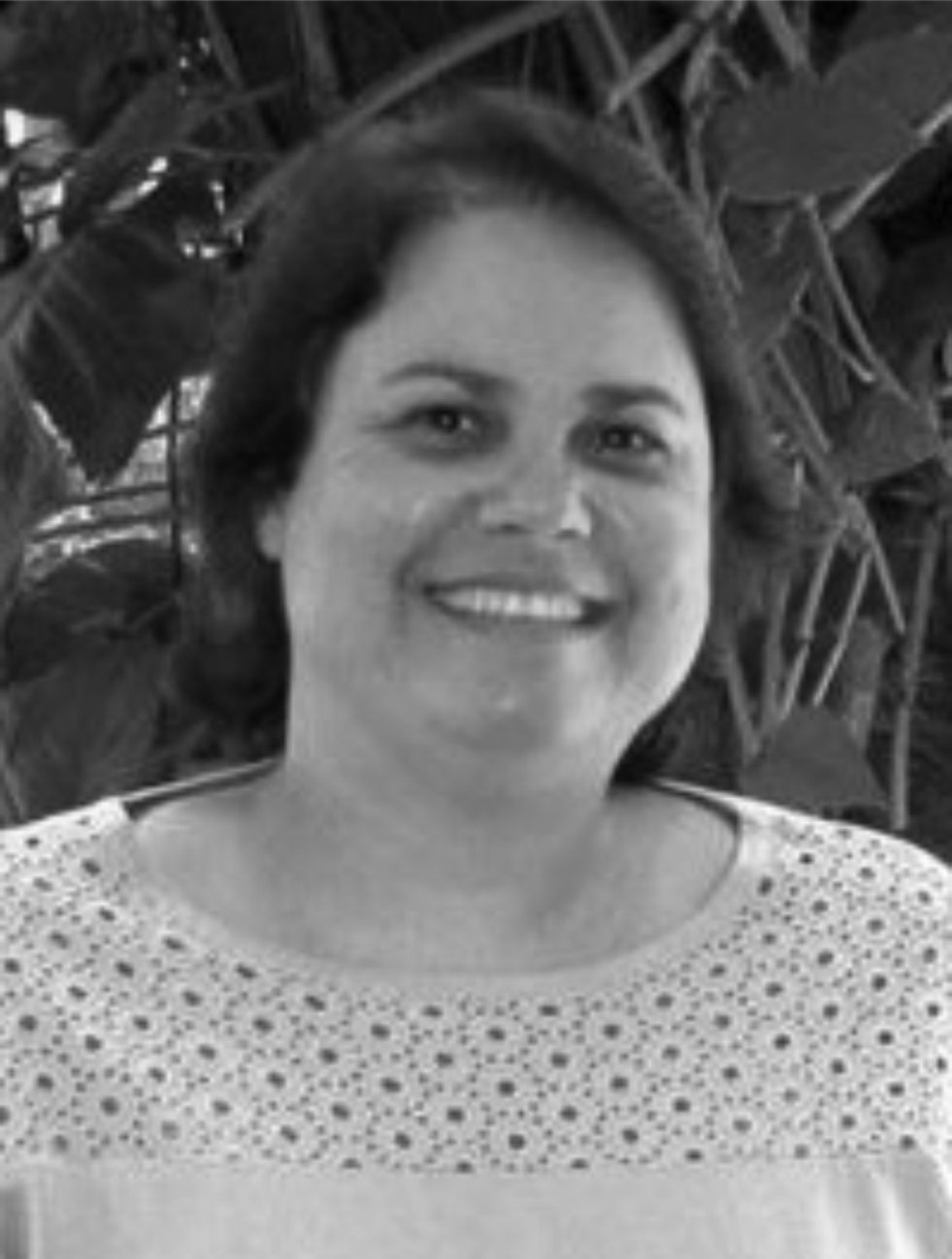}
%	\vspace*{15pt}
}%
\end{wrapfigure}
\noindent\small 
\\
\\
\\
\textbf{Agma~J.~M.~Traina} 
received the BSc and MSc degrees in Computer Science and PhD in Computational Applied Physics all from the University of São Paulo, Brazil in 1983, 1987 and 1991 respectively. She also spent a sabbatical leave as a visiting researcher at the Computer Science Department of the Carnegie Mellon University (1998-2000) working on Multimedia Databases. She is a full professor with the Mathematics and Computer Science Institute - University of São Paulo, since 2008. 
Her research interests include indexing and retrieval of complex data by content, similarity queries, data visualization, visual data mining, as well as image and video processing. Agma has been working in the integration of the results of her research lines with applications to medicine, aimed at the development of applied computational systems. She is a member of the Brazilian Computer Society, ACM and IEEE Computer Society.}

\end{document}

\maketitle

\section{Data Requirements}

To explore the ClinicalPath system, the data must contain a set of requirements to fulfill the proposed visualization. The most important are:

\begin{itemize}
    \item A list of tests containing the collection date, name of test, acronym, test result, and the patient identification; 
    
    \item A list of reference values for each test, to create the normal, low/high, and very low/high categorization; 
    
    \item A list of meta-information for each patient, such as sex and age;
    
    \item A list containing the category or type of test to create the test ordering process.
\end{itemize}

\section{Tables of Test Lists}

\noindent\begin{table}[ht]
\label{tab:vars_covid}
\caption{Test list part I.}
\centering
%\scriptsize
\begin{tabular}{ll}
    \hline
    \textbf{Acronyms} & \textbf{Description} \\
    \hline
    \textbf{covid\_iga} & detection of SARS-CoV2 by \texttt{IgA}, = 1 if true else 0\\
    \textbf{covid\_igm} & detection of SARS-CoV2 by \texttt{IgM}, = 1 if true else 0\\
    \textbf{covid\_igg} & detection of SARS-CoV2 by \texttt{IgG}, = 1 if true else 0\\
    \textbf{covid\_pcr} & detection of SARS-CoV2 by \texttt{PCR}, = 1 if true else 0\\
    \textbf{covid\_soro} & detection of SARS-CoV2 by \texttt{ECL}, = 1 if true else 0\\
    \textbf{covid\_iga\_num} & SARS-CoV2 \texttt{IgA} antibodies (AU/mL)\\
    \textbf{covid\_igm\_num} & SARS-CoV2 \texttt{IgM} antibodies (AU/mL)\\
    \textbf{covid\_igg\_num} & SARS-CoV2 \texttt{IgG} antibodies (AU/mL)\\
    \textbf{covid\_soro\_num} & SARS-CoV2 \texttt{ECL} immunoassay index (AU/mL)\\
    \hline
\end{tabular}
\end{table}

\noindent\begin{table}[ht]
\label{tab:vars_blood}
\caption{Test list part II.}
\centering
%\scriptsize
\begin{tabular}{ll}
    \hline
    \textbf{Acronyms} & \textbf{Description} \\
    \hline
    \textbf{aPTT} & activated partial thromboplastin time (s)\\
    \textbf{AT} & antithrombin activity (\%)\\
    \textbf{eGFR} & estimated glomerular filtration rate (mL/min/1.73m$^2$)\\
    \textbf{ESR} & erythrocyte sedimentation rate (mm/h)\\
    \textbf{HCT} & hematocrit [erythrocytes i.e. red blood cells] (\%)\\
    \textbf{MCH} & mean corpuscular [erythrocyte] hemoglobin (pg)\\
    \textbf{MCHC} & mean corpuscular hemoglobin concentration (g/dL)\\
    \textbf{MCV} & mean corpuscular volume (fL)\\
    \textbf{MPV} & mean platelet volume (fL)\\
    \textbf{PLT} & platelet count (/mm$^3$)\\
    \textbf{PT} & prothrombin time (s)\\
    \textbf{PT\%} & normalized prothrombin time (PT/INR) \\
    \textbf{RBC} & erythrocyte [red blood cell] count ($\times$10$^6$/mm$^3$)\\
    \textbf{RDW} & erythrocyte size distribution width (\%)\\
    \textbf{TT} & thrombin time (s)\\
    \textbf{WBC} & leucocyte [white bool cell] count (/mm$^3$)\\
    \textbf{basophil\#} & basophil count (/mm$^3$)\\
    \textbf{basophil\%} & percentage of basophils (\%)\\
    \textbf{eos\#} & eosinophil count (/mm$^3$)\\
    \textbf{eos\%} & percentage of eosinophils (\%)\\
    \textbf{lymphocyte\#} & lymphocyte count (/mm$^3$)\\
    \textbf{lymphocyte\%} & percentage of lymphocytes (\%)\\
    \textbf{monocyte\#} & monocyte count (/mm$^3$)\\
    \textbf{monocyte\%} & percentage of monocytes (\%)\\
    \textbf{neutrophil\#} & neutrophil count (/mm$^3$)\\
    \textbf{neutrophil\%} & percentage of neutrophils (\%)\\
    \hline
\end{tabular}
\end{table}

\noindent\begin{table}[ht]
\caption{\label{tab:vars_func}Test list part III.}
\centering
%\scriptsize
\begin{tabular}{ll}
    \hline
    \textbf{Acronyms} & \textbf{Description}\\
    \hline
    \textbf{ALP} & alkaline phosphatase (U/L)\\
    \textbf{ALT} & alanine [glutamic-pyruvic] transaminase  (U/L)\\
    \textbf{AST} & aspartate [glutamic-oxaloacetic] transaminase (U/L)\\
    \textbf{BILC} & conjugated bilirubin (mg/dL)\\
    \textbf{BILU} & unconjugated bilirubin (mg/dL)\\
    \textbf{Ca} & calcium (mmol/L)\\
    \textbf{Ca++} & ionized calcium (mmol/L)\\
    \textbf{Ca++F} & post-filter ionized calcium (mmol/L)\\
    \textbf{Cl-} & serum chloride (mEq/L)\\
    \textbf{CRP} & C-reactive protein (mg/dL)\\
    \textbf{cTnI} & cardiac troponin I (ng/mL)\\
    \textbf{D-D} & D dimer (ng/mL)\\
    \textbf{GGT} & gamma-glutamyl transferase (U/L)\\
    \textbf{Hb} & serum hemoglobin (g/dL)\\
    \textbf{HbA1c} & Hemoglobin A1C (mmol/mol)\\
    \textbf{HCO3-} & serum bicarbonate (mmol/L)\\
    \textbf{hs-cTnT} & high-sensitivity cardiac troponin T (ng/mL)\\
    \textbf{IL-6} & interleukin 6 (pg/mL)\\
    \textbf{IL-10} & interleukin 10 (pg/mL)\\
    \textbf{K+} & serum potassium (mEq/L)\\
    \textbf{LDH} & lactate dehydrogenase (U/L)\\
    \textbf{Na+} & serum sodium (mEq/L)\\
    \textbf{NT-proBNP} & N-terminal brain natriuretic peptide precursor (pg/mL)\\
    \textbf{PCT} & procalcitronin (ng/mL)\\
    \textbf{pH} & venous blood pH\\
    \textbf{PTH} & parathyroid hormone (pg/mL)\\
    \textbf{TBIL} & total bilirubin (mg/dL)\\
    \textbf{TNF} & tumor necrosis factor alpha (pg/mL)\\
    \textbf{TSH} & thyroid-stimulating hormone (µU/L)\\
    \textbf{albumin} & albumin (g/dL)\\
    \textbf{cholesterol} & total cholesterol (mg/dL)\\
    \textbf{creatinine} & creatinine (mg/dL)\\
    \textbf{ferritin} & ferritin (µg/L)\\
    \textbf{fibrinogen} & fibrinogen (mg/dL)\\
    \textbf{globulin} & globulin (g/dL)\\
    \textbf{glucose} & glucose (mg/dL)\\
    \textbf{protein} & total protein (g/dL)\\
    \textbf{urea} & serum urea (mg/dL)\\
    \hline
\end{tabular}
\end{table}

\clearpage
\section{Test Categorization}

\begin{table}[ht]
\centering
\caption{Test categorization and the acronyms of the test -- Part I}
\label{tab:my-table}
\begin{tabular}{ll}
\hline
\textbf{Test Category}               & \textbf{Acronyms} \\ \hline
Red Series Hemogram                  & RBC               \\
Red Series Hemogram                  & Hb                \\
Red Series Hemogram                  & HCT               \\
Red Series Hemogram                  & MCV               \\
Red Series Hemogram                  & VCM               \\
Red Series Hemogram                  & MCH               \\
Red Series Hemogram                  & MCHC              \\
Red Series Hemogram                  & RDW               \\
White Series Hemogram                & WBC               \\
White Series Hemogram                & basophil\#        \\
White Series Hemogram                & basophil\%        \\
White Series Hemogram                & eos\#             \\
White Series Hemogram                & eos\%             \\
White Series Hemogram                & lymphocyte\#      \\
White Series Hemogram                & lymphocyte\%      \\
White Series Hemogram                & monocyte\#        \\
White Series Hemogram                & monocyte\%        \\
White Series Hemogram                & neutrophil\#      \\
White Series Hemogram                & neutrophil\%      \\
Hemogram - Platelets                 & PLT               \\
Medium Platelet Volume               & MPV               \\
Liver Function / Coagulation Factors & aPTT              \\
Liver Function / Coagulation Factors & AT                \\
Liver Function / Coagulation Factors & PT                \\
Liver Function / Coagulation Factors & TT                \\
Liver Function / Coagulation Factors & ALP               \\
Liver Function / Coagulation Factors & ALT               \\
Liver Function / Coagulation Factors & AST               \\
Liver Function / Coagulation Factors & BILC              \\
Liver Function / Coagulation Factors & BILU              \\
Liver Function / Coagulation Factors & PT\%              \\
Liver Function / Coagulation Factors & D-D               \\
Liver Function / Coagulation Factors & fibrinogen        \\
Liver Function / Coagulation Factors & GGT               \\
Liver Function / Coagulation Factors & TBIL              \\
Liver Function / Coagulation Factors & albumin           \\
Liver Function                       & eGFR              \\
Liver Function                       & creatinine        \\
Liver Function                       & urea              \\ \hline
\end{tabular}%
\end{table}

\begin{table}[ht]
\centering
\caption{Test categorization and the acronyms of the test -- Part II}
\label{tab:my-table_part2}
\begin{tabular}{ll}
\hline
\textbf{Test Category}  & \textbf{Acronyms} \\ \hline
Ion Evaluation          & Ca                \\
Ion Evaluation          & Ca++              \\
Ion Evaluation          & Ca++F             \\
Ion Evaluation          & Cl-               \\
Ion Evaluation          & HCO3-             \\
Ion Evaluation          & K+                \\
Ion Evaluation          & Na+               \\
Ion Evaluation          & pH                \\
Cardio Evaluation       & cTnI              \\
Cardio Evaluation       & cTnT              \\
Cardio Evaluation       & NT-proBNP         \\
Inflammatory Evaluation & CRP               \\
Inflammatory Evaluation & PCT               \\
Inflammatory Evaluation & ESR               \\
Inflammatory Evaluation & globulin          \\
Inflammatory Evaluation & IL-6              \\
Inflammatory Evaluation & IL-10             \\
Inflammatory Evaluation & TNFa              \\
Inflammatory Evaluation & LDH               \\
Endocrine Evaluation    & glucose           \\
Endocrine Evaluation    & HbA1c             \\
Endocrine Evaluation    & TSH               \\
Endocrine Evaluation    & PTH               \\
General Evaluation      & cholesterol       \\
General Evaluation      & ferritin          \\
General Evaluation      & protein           \\
COVID                   & covid\_pcr        \\
COVID                   & covid\_iga        \\
COVID                   & covid\_soro       \\
COVID                   & covid\_igg        \\
COVID                   & covid\_igm        \\ \hline
\end{tabular}%
\end{table}

\newpage
\section{Questionnaire from User Evaluation}

\begin{table*}[ht]
\centering
\caption{Background and experience questions.}
\label{tab:background_experience}
\begin{tabular}{lll}
\cline{1-2}
\textbf{Acronyms} & \textbf{Background and experience questions}                                  &  \\ \cline{1-2}
\textbf{BQ1}      & How old are you?                                                              &  \\ \\
\textbf{BQ2}      & What is your gender?                                                          &  \\ \\
\textbf{BQ3}      & How many years of experience do you have in the field of medicine in general? &  \\ \\
\textbf{BQ4}      & Do you have any medical specialization? If yes, which one?                    &  \\ \cline{1-2}
\end{tabular}%
\end{table*}

\begin{table*}[ht]
\centering
\caption{Three basic questions not requiring medical knowledge. Correct answers are marked in bold.}
\label{tab:simple_questions}
\begin{tabular}{ll}
\hline
\textbf{Acronyms} & \textbf{Question description}                                                                                                                                                                                                                                   \\ \hline
\textbf{SQD}               & \begin{tabular}[c]{@{}l@{}}Open the ClinicalPath system and choose patient 268138 to view the Clinical path. \\ In the red series blood count category, select the Serum Hemoglobin (HB) test \\ and answer the following questions:\end{tabular}                 \\ \\
\textbf{SQ1}               & \begin{tabular}[c]{@{}l@{}}On 11/13/2019, what was the result of this test?\\ Choices: 13,1; \textbf{10,6}; 18,1; 19; 9,3.\end{tabular}                                                                                                                                  \\ \\
\textbf{SQ2}               & \begin{tabular}[c]{@{}l@{}}On 01/02/2020, in which value range was the result of this test in agreement \\ with the visualization?\\ Choices: Very low; low; \textbf{normal}; high; very high.\end{tabular}                                                              \\ \\
\textbf{SQ3}               & \begin{tabular}[c]{@{}l@{}}Deselect the HB test and select the HCM test. According to this patient's \\ entire history, the result of this test was always:\\ Choices: \textbf{Normal}; abnormal; changes over time; this patient did not perform this test.\end{tabular} \\ \hline
\end{tabular}%
\end{table*}

\begin{table*}[ht]
\centering
\caption{Five advanced questions containing a fictional patient's medical history (four multiple-choice and one open question). The possible answers for AQ1 -- AQ4 are: Yes; no; I don't know. The correct answers were Yes, Yes, No, Yes for AQ1, AQ2, AQ3, and AQ4 respectively.}
\label{tab:adv_questions}
\begin{tabular}{ll}
\hline
\textbf{Acronyms} & \textbf{Question description}                                                                                                   \\ \hline
\textbf{AQD}               & \begin{tabular}[c]{@{}l@{}}In the ClinicalPath system, go to "Open" in the upper left corner and select\\  patient 1591522 and read the clinical history. \\ \textit{Clinical history of the patient}: Patient MRS, 36 years old, puerperal, in the \\ 15th Postoperative after cesarean, at the present date (10/01/2020) comes to the emergency \\ room with complaints of fatigue, feelings of fainting or pressure drop, reporting \\ skin sensation hot, and have observed slight red vaginal bleeding. On physical \\ examination: eupneic in Ambient Air, blood pressure: 85x60, respiratory frequency: 26, \\ heart rate: 110, St O2: 94\%, axillary temperature: 99.5 fahrenheit.  \\ Abdomen: flaccid, slightly painful on palpation, especially \\ in the lower quadrants, surgical wound reddened at the edges, with a small area \\ of drainage of seropurulent secretion. Without too many particularities.\\ Read only the tests requested in the questions to solve them. Based on the \\ the history presented, review the tests requested in the system, and answer:\end{tabular} \\ \\
\textbf{AQ1}               & \begin{tabular}[c]{@{}l@{}}When evaluating the blood counts from 10/01/2020 to 10/05/2020, is it possible \\ to infer that there is a possibility of late puerperal hemorrhage, with significant \\ drops in Hb and Ht? (late puerperium: 11th day to 42nd day postpartum)\end{tabular}                                                                                                                                                                                                                                                                                                                                                                                                                                                                                                                                                                                                                                                                                                                         \\ \\
\textbf{AQ2}               & \begin{tabular}[c]{@{}l@{}}Analyzing the white blood cell count (WBC), do you believe it is possible to be facing\\  an infection resulting from the assistance offered in a hospital environment?\end{tabular}                                                                                                                                                                                                                                                                                                                                                                                                                                                                                                                                                                                                                                                                                                                                                                                                       \\ \\
\textbf{AQ3}               & \begin{tabular}[c]{@{}l@{}}Evaluate the results of PLT, aPTT, and PT\% in the mentioned period, do\\ you agree that, in case there is active bleeding or important inflammatory\\ disorder, these did not impact the clotting mechanisms of this patient?\end{tabular}                                                                                                                                                                                                                                                                                                                                                                                                                                                                                                                                                                                                                                                                                                                                      \\ \\
\textbf{AQ4}               & \begin{tabular}[c]{@{}l@{}}Analyzing the results of Urea and Creatinine in the period from 10/01/2020 \\ to 11/24/2020, knowing that antibiotic therapy was used, is it plausible to \\ say that there was an acute renal failure and that it was solved?\end{tabular}                                                                                                                                                                                                                                                                                                                                                                                                                                                                                                                                                                                                                                                                                                                                          \\ \\

\textbf{AQ5}               & \begin{tabular}[c]{@{}l@{}}Based on the tests and the history presented, do you think it is possible \\ to create other diagnostic hypotheses or identify relevant changes for  this \\ patient that are not included in the previous questions? If yes, name some \\ that you consider relevant, and discuss how did you found them. \end{tabular}                                                                                                                                                                                                                                                                                                                                                                                                                                                                                                                                                                                                                                                                                                                                          \\ \hline
\end{tabular}%
\end{table*}

\begin{table*}[ht]
\centering
\caption{Likert-scale-based questions to evaluate the user preference of the system, along with an open question to justify the choices. The five options of the 5-Likert-scale are: Strongly Disagree, Disagree, I don't know, Agree and Strongly Agree.}
\label{tab:likert_scale}
\begin{tabular}{ll}
\hline
\textbf{Acronyms} & \textbf{Question description}                                                                                                                                           \\ \hline
\textbf{LQD}      & \begin{tabular}[c]{@{}l@{}}According to your opinion about the system, rate using a \\ 5-Likert-scale according to the level of agreement of the sentence:\end{tabular} \\ \\
\textbf{LQ1}      & The interface and usability of ClinicalPath is intuitive.                                                                                                               \\ \\
\textbf{LQ2}      & ClinicalPath is efficient and would optimize my time.                                                                                                                   \\ \\
\textbf{LQ3}      & ClinicalPath is easy to use.                                                                                                                                            \\ \\
\textbf{LQ4}      & It's easy to learn how to use ClinicalPath.                                                                                                                             \\ \\
\textbf{LQ5}      & \begin{tabular}[c]{@{}l@{}}Leave a comment on your perception of the ClinicalPath system \\ with regard to the above items.\end{tabular}                                \\ \hline
\end{tabular}%
\end{table*}

\begin{table*}[ht]
\centering
\caption{A mix of multiple choices and open questions to collect the users' feedback about the system. }
\label{tab:feedback}
\begin{tabular}{ll}
\hline
\textbf{Acronyms} & \textbf{Question description}                                                                                                                                            \\ \hline
\textbf{FQD}       & \begin{tabular}[c]{@{}l@{}}According to your opinion, answer the questions \\ about your feedback of the system.\end{tabular}                                            \\ \\
\textbf{FQ1}       & \begin{tabular}[c]{@{}l@{}}Have you ever tried to do analyzes similar to those \\ performed on the tasks described in the experiment? \\ (Yes or No choice)\end{tabular} \\ \\
\textbf{FQ2}       & \begin{tabular}[c]{@{}l@{}}Did you gain insights or come to new conclusions from \\ your exploration of the system? (Yes or No choice)\end{tabular}                      \\ \\
\textbf{FQ3}       & \begin{tabular}[c]{@{}l@{}}In your opinion, what are the most useful visual aids \\ offered by ClinicalPath? (Open question)\end{tabular}                                \\ \\
\textbf{FQ4}       & \begin{tabular}[c]{@{}l@{}}What other visual aids do you think could be useful \\ if incorporated into the ClinicalPath system? (Open question)\end{tabular}             \\ \hline
\end{tabular}%
\end{table*}